\documentclass[aps,prx,twocolumn,superscriptaddress,groupedaddress]{revtex4}
\pdfoutput=1
\usepackage{graphicx}  % needed for figures
\usepackage{dcolumn}   % needed for some tables
\usepackage{bm}        % for math
\usepackage{amssymb}   % for math
\usepackage{amsmath}   % for math
\usepackage{mathtools} % for math
\usepackage{color}
\usepackage{xcolor}
\usepackage{ulem}

\usepackage[colorlinks,linkcolor=blue,urlcolor=blue,citecolor=blue]{hyperref}

 \newcommand{\aref}[1]{Appendix~\ref{#1}}%

\newcommand\be{\begin{equation}} 
\newcommand\ee{\end{equation}} 
\begin{document}

\title{Entropy production in systems with unidirectional transitions}
\author{D. M. Busiello}
\thanks{These authors equally contributed to this work}
\affiliation{Ecole Polytechnique F\'ed\'erale de Lausanne (EPFL), Institute of Physics Laboratory of Statistical Biophysics, 1015 Lausanne, Switzerland}
\author{D. Gupta}
\thanks{These authors equally contributed to this work}
\affiliation{Dipartimento di Fisica `G. Galilei', INFN, Universit\'a di Padova, Via Marzolo 8, 35131 Padova, Italy}
\author{A. Maritan}
\affiliation{Dipartimento di Fisica `G. Galilei', INFN, Universit\'a di Padova, Via Marzolo 8, 35131 Padova, Italy}

\date{\today}

\begin{abstract}
The entropy production is one of the most essential features for systems operating out of equilibrium. The formulation for discrete-state systems goes back to the celebrated Schnakenberg's work and hitherto can be carried out when for each transition between two states also the reverse one is allowed. Nevertheless, several physical systems may exhibit a mixture of both unidirectional and bidirectional transitions, and how to properly define the entropy production in this case is still an open question. Here, we present a solution to such a challenging problem. The average entropy production can be consistently defined, employing a mapping that preserves the average fluxes, and its physical interpretation is provided. We describe a class of stochastic systems composed of unidirectional links forming cycles and detailed-balanced bidirectional links, showing that they behave in a pseudo-deterministic fashion. This approach is applied to a system with time-dependent stochastic resetting. Our framework is consistent with thermodynamics and leads to some intriguing observations on the relation between the arrow of time and the average entropy production for resetting events.
\end{abstract}

%\noindent{\bf Keywords:} {}

\maketitle

%%% To make the table of contents %%%%%
%\noindent\rule{\hsize}{2pt}
%\tableofcontents
%\noindent\rule{\hsize}{2pt}
\markboth{Entropy production in systems with unidirectional transitions}{}

\section{Introduction}
\label{intro}
Almost all natural systems operate out of equilibrium, producing entropy in their surroundings. A general situation for a system to show a non-equilibrium dynamics, and to eventually achieve a non-equilibrium stationary state, is to be in contact with different reservoirs (e.g., of energy, volume, matter) not necessarily in a mutual equilibrium \cite{schn}. The starting point to model a system in this situation is to perform a coarse-graining procedure on some degrees of freedom, ending up with a set of equations describing the system in terms of probabilities and transition rates.

Given a dynamical description, the most interesting features of a system out of equilibrium arise from the study of its thermodynamic properties. In particular, the fingerprint of a non-equilibrium condition is the entropy production. It has been widely investigated and found to be of paramount importance in estimating the accuracy of relevant observables of natural systems from several perspectives \cite{dechant,barato,horo,jarz,Esposito-CG}. Moreover, the entropy production plays a leading role in controlling the linear response to external stimuli \cite{baiesi} and allows determining the non-equilibrium stationary state of the system, in close to equilibrium conditions \cite{prigogine, vanden}.

Consider a discrete-state Markovian system whose dynamics can be described by a Master Equation \cite{schn}. In formulas:
\begin{align}
\dfrac{dp_i{\color{black}(t)}}{dt}=\sum_{j=1}^{N} [W_{j\to i}p_j(t)-W_{i\to j}p_i(t)],
\label{mp}
\end{align}
where $p_i(t)$ indicates the probability to be in the state $i$ at time $t$, $W_{i\to j}$ the transition rate to pass from the state $i$ to the state $j$, that can also depend on time, and $N$ the total number of states. The entropy production of such a system can be estimated using the Schnakenberg's formula \cite{schn}:
\begin{equation}
\dot{S}_{tot} = \frac{1}{2} \sum_{i,j} \left(W_{j\to i} p_j - W_{i\to j} p_i \right) \log\dfrac{W_{j\to i} p_j}{W_{i\to j} p_i} \geq 0,
\label{tot-ent-1}
\end{equation}
 where $\dot{S}_{tot} =0$ corresponds to the system in the thermodynamic equilibrium.

The quantity $\dot{S}_{tot}$ can be further splitted into an environment contribution, $\dot{S}_{env}$, and a system one, $\dot{S}_{sys}$, this latter vanishing at the stationary state, as follows:
\begin{align}
\dot{S}_{env} &= \frac{1}{2} \sum_{i,j} \left( W_{j\to i} p_j - W_{i\to j} p_i \right) \log\dfrac{W_{j\to i}}{W_{i\to j}}, \label{eqr} \\
\dot{S}_{sys} &= \frac{1}{2} \sum_{i,j} \left( W_{j\to i} p_j - W_{i\to j} p_i \right) \log\dfrac{p_j}{p_i}, \label{eqs}
\end{align}
 where $S_{sys}=-\sum_i p_i \log p_i$.
The usual assumption is that if $W_{i\to j} \neq 0$, so it is $W_{j\to i}$ so that \eqref{tot-ent-1} and \eqref{eqr} are well defined. Nevertheless, there is a vast class of systems for which this assumption is not satisfied. For example, total asymmetric simple exclusion process (TASEP) involves the transition only in one direction while respecting the exclusion mechanism \cite{Ronald,saha}. {\it Unidirectional} jumps are also used  to model biological enzymatic reactions \cite{Ting}. The lack of microscopic reversibility can be seen in uniform sheared granular matter and driven inelastic Lorentz-gas \cite{Chong}. Some other examples include directed percolation \cite{Masaki}, spontaneous decay of an excited atom \cite{rahav}, and stochastic resetting {\color{black}\cite{search1, reviewresetting}}. The onset of this peculiar situation can also be due to the presence of different processes driving the system out of equilibrium. Once we identify each single process, a transition between $i$ and $j$ can be unidirectional for one of them, but possible along a different pathway. The existence of multiple channels allows for a microscopic interpretation of the local detailed balance \cite{Paolo}. However, as one can easily see from \eqref{tot-ent-1}, the entropy production cannot be defined when there are some unidirectional transitions \cite{hinric,saha}. 

In this direction, several efforts have been made in understanding and describing this vast class of systems. For example, Ben-Avraham {\it et al.} \cite{ben2011entropy} tackled the problem employing a coarse-graining of the sampling time and derived the entropy production that diverges as the sampling time approaches zero, i.e., counting each transition. Similar results on the temporal coarse-graining can also be seen in \cite{hinric}. In contrast to these studies, Murashita {\it et al.} \cite{Murashita} identified the reasoning behind this divergence at a trajectory level by separating regular and irregular trajectories, and obtaining a correction to the integral fluctuation theorem. In contrast,  some fluctuation theorems were derived by Ohkubo \cite{ohkubo}, using a Bayesian approach and defining a posterior probability to describe the reverse process.
%gave an alternative technique to understand the various known fluctuation theorems using the posterior probability instead of probability of trajectory in a reverse process. 
Recently,  Rahav {\it et al.} \cite{rahav} studied a fluctuating quantity similar (but not equal) to the entropy production in a system with unidirectional transitions, finding that it obeys an integral fluctuation theorem.
%which depends on the forward trajectory and the reverse one of the so called an {\it auxiliary dynamics} and this quantity is shown to obey integral fluctuation theorem.  
However, none of these methods can be useful in defining the entropy production without observing a divergence absent in experiments \cite{lacasta,Ronald}. Then, we present an argument to evaluate the average entropy production even in presence of unidirectional transitions thus generalizing Schnakenberg's original work. 

We apply our derivations to a peculiar situation in which detailed balance is satisfied on bidirectional transitions, leading to a pseudo-deterministic behavior, i.e., on average, a null entropy production emerging from unidirectional cycles.

Further, we consider the well-known problem of resetting. This is a process involving a sudden transition to a single preselected state or region of the system. It has attracted attention in optimization problems \cite{search1,search2,search3}, proofreading \cite{proof1,proof2}, population dynamics \cite{popdyn} and information theory \cite{info}. We derive the results presented in \cite{resetting} in our general framework. 

Finally, we extend the study of systems with resetting to cases in which this latter mechanism is periodic in time. On the one hand, the time-periodic driving has been recently studied as a tool to mimic desired non-equilibrium features (e.g. stationary distribution, entropy production, fluxes) of biological systems in which detailed balance does not hold \cite{browne,hern,astumian,raz-subasi,busiello-raz}. Since some of them may exhibit a resetting mechanism or, more in general, unidirectional transitions, then a way to estimate the entropy production both in steady and time-periodic conditions is of paramount importance to efficiently engineering artificial molecular machines. On the other hand, the paradigm of time-periodic resetting presents a much more wide phenomenology than its stationary counterpart. Indeed, we show that it allows switching between a behavior where information is erased from the system to one where, instead, it is added. Interestingly, a signature of the arrow of time emerges by inspecting the average resetting entropy production. Even if presented on a simple toy model, this could serve as a starting point to elucidate the response of biological systems to time-periodic stimuli \cite{bennett, mettetal,tu,timeperiodic}.

The plan of this paper is as follows. In Sec. \ref{me-ft}, we give a general discussion on a system involving both bidirectional and unidirectional transitions, and identify the total entropy production for a general system obeying Eq. \eqref{mp}. Sec. \ref{dyna} presents the mapping of unidirectional links to bidirectional ones. Here we show that under suitable limiting procedure of the transition rates of these bidirectional links, one is able to derive the total entropy production identified in Sec. \ref{me-ft}. {\color{black}An} interpretation of this limit is given in Subsec. \ref{inter}. Using the entropy production along a single trajectory of a non-equilibrium ensemble, the results shown in Sec. \ref{dyna} are \textcolor{black}{reobtained} in Sec. \ref{traj}.  Moreover, in Subsec. \ref{PDB}, we consider some examples to better understand the entropy production along a stochastic trajectory when the system has a mixture of both unidirectional and bidirectional links. Entropy production for a system under stochastic resetting mechanism is \textcolor{black}{treated} in Sec. \ref{st-rt}. Using the general theory we study the entropy production using a constant resetting rate (Subsec. \ref{constt-res}) and the time-periodic variation of the the resetting rate (Subsec. \ref{timeresett}) in a model system. In Sec. \ref{conc}, we conclude our paper. A detailed derivation for the averaging of entropy on a stochastic trajectory over the path probability is shown in \aref{path-prob}. A relation between the rate of total entropy production and the total entropy production along a stochastic trajectory is shown in \aref{proof}.

\section{Master Equation with unidirectional transitions}
\label{me-ft}
\begin{figure}[h]\center
    \includegraphics[width=8 cm]{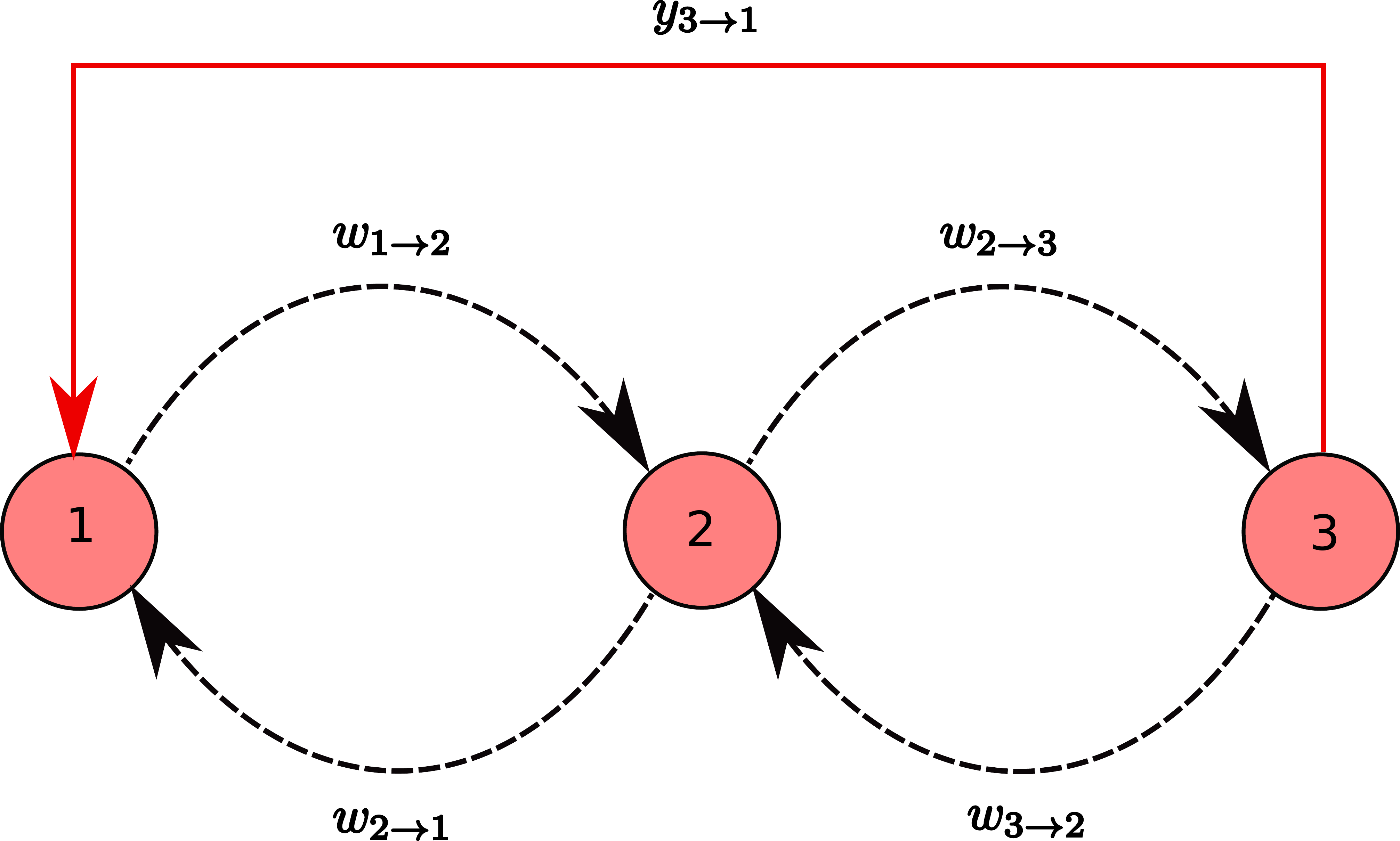}
    \caption{\label{fig-1} A three state system in which states are labeled by $1,2,3$ is shown. Each black dashed link represents the transitions from one state to the other and has a corresponding conjugate link which describes the reverse transition. On the other hand, there is no reverse transition with respect to that of the red solid link.}
\end{figure}

We consider a discrete state Markov system as described by Eq. \eqref{mp}. The system follows a stochastic dynamics and makes transitions from one state to the other. For a selected pair of states (e.g., $i,j$), either (i) the system jumps from a state $i$ to a state $j$ with a transition rate {\color{black}$W_{i \to j} \equiv w_{i\to j}>0$} and there exists a corresponding reverse transition with a rate {\color{black}$W_{j \to i} \equiv w_{j\to i}>0$} (for example, see black dashed links in Fig. \ref{fig-1}) or (ii) the system makes a transition from a state $i$ to a state $j$ with a transition rate {\color{black}$W_{i \to j} \equiv y_{i\to j}>0$} in the absence of a conjugate transition, i.e., {\color{black}$W_{j \to i} \equiv y_{j\to i}=0$}, (for example, see the red link in Fig. \ref{fig-1}). The former process is referred to as bidirectional process whereas the latter is called unidirectional process.

The dynamics of the system is described by a continuous-time Master Equation:
\begin{align}
\dfrac{dp_i}{dt}=\sum_{j=1}^{N} (w_{j\to i}p_j-w_{i\to j}p_i)+\sum_{j=1}^{N}(y_{j\to i}p_j-y_{i\to j}p_i),
\label{dyn-first}
\end{align}  
where on the right hand side, the first term describes the situation (i), and the second term corresponds to the situation (ii) in which the summation is performed over all the couple of transition rates for which the reverse transitions are not allowed, i.e., either $(y_{i \to j}=0, y_{j \to i}\neq0)$ or $(y_{j \to i}=0, y_{i \to j}\neq0)$. {\color{black}From now on we will not always explicitly write the time dependence in $p_i$ and in the transition rates.}

The average entropy of the system is given by \textcolor{black}{\cite{schn}} (as it is common in this context we use temperature units so that the Boltzmann constant {\color{black} can be set} equal to $1$)
\begin{align}
S_{sys}=-\sum_{i=1}^{N}p_i \log {p_i},
\end{align}
where $p_i(t)$ is the solution of the Master Equation \eqref{dyn-first} subject to the initial condition $p_i(0)$.

Differentiating the above equation with respect to time $t$ and substituting \eqref{dyn-first} yields the system entropy production $\dot S_{sys}$:
\begin{align}
\dot S_{sys}=\sum_{i,j} w_{j\to i}p_j \log\dfrac{p_j}{p_i}+\sum_{i,j} y_{j\to i}p_j \log\dfrac{p_j}{p_i}, \label{eq7}
\end{align}  
where the dot represents the derivative with respect to time.  Notice that $\dot S_{sys}$ is a finite quantity irrespective  of the fact that some of the transitions are not allowed.  The above equation can be rewritten as follows 
{\color{black}
\begin{eqnarray}
\dot S_{sys} &=& \overbrace{\sum_{i,j} w_{j\to i}p_j \log\dfrac{w_{j\to i}p_j}{w_{i\to j}p_i}}^{\dot S_{tot}^{(R)}}- \overbrace{\sum_{i,j} w_{j\to i}p_j \log\dfrac{w_{j\to i}}{w_{i\to j}}}^{\dot S_{env}^{(R)}} + \nonumber \\
&\;& + \underbrace{\sum_{i,j} y_{j\to i}p_j \log\dfrac{p_j}{p_i}}_{-\dot{S}^{(\rm u)}},
\label{parts}
\end{eqnarray}}
where the first and second term on the right hand side, respectively, are recognized as the total entropy production $\dot S^{(R)}_{tot}$ and the environment entropy production $\dot S^{(R)}_{env}$ for the bidirectional processes only.  The superscript $R$ refers only to the fact that the corresponding quantities contain only bidirectional transition rates, however, the information regarding both unidirectional and bidirectional links is encoded in $p$'s. 

{\it Remark 1}: both $\dot{S}_{tot}^{(R)}$ and $\dot{S}_{env}^{(R)}$ depend on the whole set of transition rates, $\{w\}$ and $\{y\}$, through the $p$'s that satisfy Eq. \eqref{dyn-first}.

{\it Remark 2}: the contribution of the unidirectional links (i.e., last term on the right hand side) cannot be splitted into a total and an environment entropy production, since both terms will formally lead to divergent (and unphysical) quantities.

{\color{black}We name ``unidirectional entropy production'', $\dot{S}^{(\rm u)}$, the last term on the r.h.s. in Eq. \eqref{parts} explicitly depending on unidirectional transition rates. This extra contribution has to be investigated case by case to understand its physical origin. When, for example, there is an external process inducing unidirectional transitions whose entropy production is negligible, $\dot{S}^{(u)}$ can be understood as an additional contribution to the environment entropy production. In the next section we corroborate the latter physical interpretation with an explicit example.}

%We claim that $\dot{S}_{tot}^{(R)}$ in Eq. \eqref{parts} should be identified as the total entropy production{\color{black}, $\dot S_{tot}$, associated with Eq. \eqref{dyn-first}}. This implies that the sum of the last two terms in the same equation is equal to $-\dot{S}_{env}$, i.e., the entropy production associated to the environment, with a minus sign. In the next section, we will present an argument supporting our hypothesis.

\section{Dynamics-preserving irreversible entropy production}
\label{dyna}
\begin{figure*}[t]\center
    \includegraphics[width=15 cm]{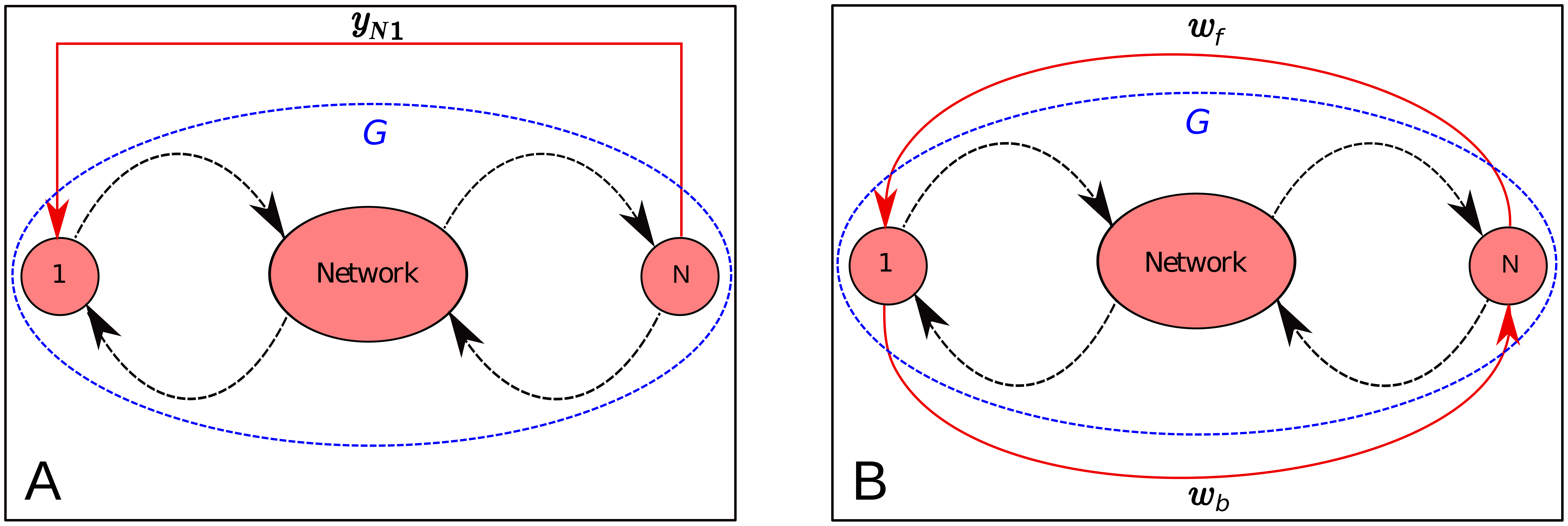}
    \caption{\label{fig-2} A) A Master Equation system with $N$ states, defined by a network of transitions $\textit{G}$, with an external unidirectional link $y_{N1}$. B) A fictitious system with the same dynamics of the one in (A), where the only difference is the presence of bidirectional transitions between the states $1$ and $N$, as indicated, whose strengths are determined through Eq. \eqref{fluxes}}.
\end{figure*}
Let us consider the discrete-state system in Fig. \ref{fig-2}A, say $(A)$. All the transition rates present in the internal network $\textit{G}$ are bidirectional. The only unidirectional link is the external one, $y_{N1}:=y_{N\to1}$. This system is physically equivalent to the one depicted in Fig. \ref{fig-2}B, say $(B)$, as long as the flux flowing through the unidirectional link (A) is equal to the corresponding bidirectional links (B) for all time $t$. In formulas, this condition is
\begin{align}
y_{N1} p_N(t) = w_f(t) p_N(t) - w_b p_1(t) \equiv J(t),
\label{fluxes}
\end{align}
where we have set $w_b$ to be time-independent. \textcolor{black}{In the above equation \eqref{fluxes}, in order to obtain $w_f(t)$, we substitute $p_1(t)$ and $p_N(t)$, the solutions obtained from \eqref{dyn-first} for a given initial condition $[p_1(0),p_2(0),p_3(0),\dots,p_N(0)]$ and $w_b$.} Notice that $w_b \neq 0$, since $w_b = 0$ implies $w_f = y_{N \to 1}$, corresponding to the original situation depicted in $(A)$. With this condition the dynamics of the two systems, $(A)$ and $(B)$, leads to the same average entropy production {\color{black} as in Eq. \eqref{eq7}}. 

Let us first evaluate the environment entropy production for the system $(B)$, $\dot{S}_{env}^{(B)}$. By definition, using Eq. \eqref{tot-ent-1} we have:
\begin{align}
\dot{S}_{env}^{(B)} = \dot{S}_{env}^{(\textit{G})} + J \log\left(\frac{w_f}{w_b}\right),
\label{ents}
\end{align}
where the first term is the environment entropy production due to the internal network $\textit{G}$. Using Eq. \eqref{fluxes}, we can express $w_f$ as a function of $y_{N1}$ and $w_b$ in Eq. \eqref{ents}, obtaining:
\begin{align}
\dot{S}_{env}^{(B)} = \dot{S}_{env}^{(\textit{G})} + J \log\left(\frac{y_{N1} p_N + w_b p_1}{w_b p_N}\right),
\label{env}
\end{align}
where $p_i$ indicates the probability to be in the state $i$, due to all the transition rates, both the bidirectional and the unidirectional ones, as obtained from the solution of Eq. \eqref{dyn-first}.

In order to eliminate the information about the fictitious transition rate $w_b$, we can take the limit $w_b \rightarrow +\infty$. {\color{black}The physical interpretation of this limit corresponds to the assumption that the external process inducing unidirectional transitions in the system, if any, has a negligible entropy production.} In the next subsection we will discuss {\color{black} further this point.} We get the following equation:
\begin{align}
{\color{black}\dot{S}_{env}^{(A)} \equiv \dot{S}_{env}^{(B)}\big|_{w_b \to \infty}} = \dot{S}_{env}^{(\textit{G})}-J \log\left(\frac{p_N}{p_1}\right).
\label{envlimit}
\end{align}
Thus we have obtained a (non-divergent) contribution to the environment entropy production due to the unidirectional links only, which we identify with a part of the environment entropy production {\color{black}of the system of interest $(A)$}, $\dot{S}_{env}^{(A)}$. Notice that the environment entropy production $\dot S_{env}^{(A)}$ explicitly depends on the transition rate of the unidirectional link just through the probability flux  $J = y_{N1} p_N$.

The generalization to the case of multiple unidirectional links is straightforward and proceed along the same line. In fact, we can imagine a fictitious system where each unidirectional link $y_{i\to j}$ is replaced by a couple of links $w_{f,i\to j}$ and $w_{b,j\to i}$, properly tuned according to Eq. \eqref{fluxes}. Then, after the evaluation of the entropy production, we can set all the $w_b$'s to infinity as in the above example.

Rewriting the total entropy production, from Eq. \eqref{parts}, we obtain:
\begin{gather}
\overbrace{ \sum_{i,j} w_{j\to i}p_j \log\dfrac{w_{j\to i}p_j}{w_{i\to j}p_i}}^{{\color{black}\dot{S}^{(R)}_{tot}}}= \nonumber \\
= \dot{S}_{sys} + \overbrace{\dot{S}^{(R)}_{env} + \sum_{i,j} y_{j\to i}p_j \log\dfrac{p_i}{p_j}}^{\dot{S}_{env}}~\equiv~\dot{S}_{tot}.
\label{totalEP}
\end{gather}
We stress again that the probabilities $p$'s depend on the whole network through Eq. \eqref{dyn-first}, and so also $\dot{S}_{tot}^{(R)}$ and  $\dot{S}_{env}^{(R)}$. Only the environment entropy production depends explicitly on the unidirectional links $y$'s. Furthermore, $\dot{S}_{tot}$ does not depend on the choice of $w_b$.

\begin{figure*}\center
\includegraphics[width=1.8 \columnwidth]{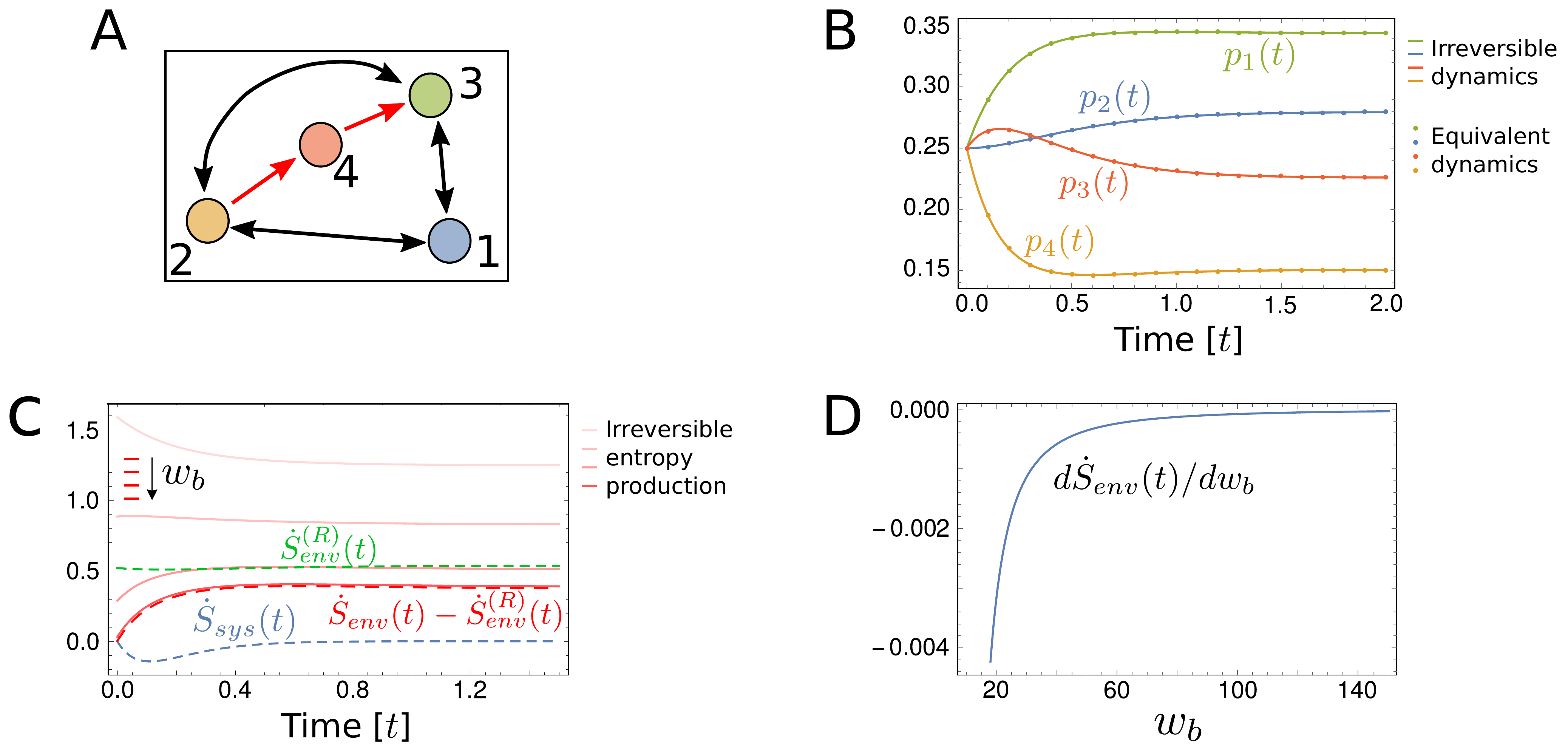}
\caption{\label{fig-3} A) A simple network with four nodes and two unidirectional transitions (shown in red) is considered as an example. B) The dynamics of the system in (A) is compared with the dynamics of the equivalent network where each unidirectional link has been mapped into two bidirectional links, as explained in Fig. \ref{fig-2}. The equivalence is valid for any choice of the parameters. In the figure we set: $w_{1 \to 2} = w_{2 \to 3} = w_{4 \to 3}=w_{3\to 1} = 2$, $w_{2 \to 4} = 3$, {\color{black} all the} others equal to unity, and the initial state $p_i(0)=0.25$ for $i=\{1,2,3,4\}$. C) The dashed blue line is the system entropy production {\color{black} as a function of time}, which goes to zero as the system approaches stationarity. The green dashed line is the evironment entropy production only due to the bidirectional links {\color{black}($\dot S^{(R)}_{env}$)}. The solid {\color{black}red} lines represent the enviornment entropy production associated only to the unidirectional links {\color{black}modified as in (the last term of) Eq. \eqref{env}, introducing equivalent bidirectional links for each unidirectional link, as shown in \ref{fig-2}}. The color becomes darker as $w_b$ increases, up to a limiting curve (red dashed line) corresponding to the last term in Eq.~\eqref{envlimit}, generalized to the case of two unidirectional links (see Eq. \eqref{totalEP}). For sake of simplicity we used just one $w_b$, equal for both the unidirectional links. D) The derivative of the environment entropy production is shown, highlighting that {\color{black}it approaches} zero as $w_b$ increases.}
\end{figure*}

We present the above results for an illustrative simple example. Consider the network consisting of four nodes depicted in Fig.~\ref{fig-3}A. In Fig. \ref{fig-3}B, we show the equivalence between the dynamics of the network with the unidirectional links (solid lines) and the one with the bidirectional links (dots) for all times, as discussed in \eqref{fluxes}, for a particular choice of transition rates. In Fig.~\ref{fig-3}C, all the components of the entropy production are studied. In particular, the system entropy production tends to zero as the system approaches the stationary state. The environment entropy productions due to bidirectional  and unidirectional links also reach a stationary value for large times. Moreover, the latter contribution (the last term in \eqref{env})  is plotted for different values of $w_b$, exhibiting its convergence to a minimum value for all the times as $w_b$ goes to infinite. In the Fig.~\ref{fig-3}D, we show that with an increasing value of $w_b$, the derivative of  $\dot S_{env}$ with respect to $w_b$ tends to zero, i.e.,  $\dot S_{env}$ becomes independent of it. Thus, the environment entropy production for the modified system, the one with $w_f$'s and $w_b$'s, is bounded from below by the environment entropy production obtained in the limit $w_b \to +\infty$, i.e., the one that we have identified as the environment entropy production of the original system.

\subsection{Interpretation of the $w_b \to +\infty$ limit}
\label{inter}
The physical interpretation of this limit,  as discussed in \cite{busielloPRE}, can be found in parallel to an electrical circuit made only by conductances, $w_{ij}$. The injection of a net current $J$ in a node $k$, and the ejection of the same current from another node $l$ can be introduced into the dynamics as a unidirectional link, $y_{l \to k}$. Analogously, the production of this current can also be seen as the effect of an external battery, producing a difference of potential $\Delta V = J R_{eq}$, where $R_{eq}$ is the equivalent resistance, which can be estimated as a combination of the bidirectional transition rates \cite{busielloPRE}. It can be shown that, if we map this battery into bidirectional links, with transition rates $w_f(t)$ and $w_b$, producing the same flux, the entropy production of the system always contains the energy dissipation of the electric circuit, and an additional contribution proportional to $1/w_b$. Letting $w_b$ go to infinity corresponds to a circuit where the current is generated by a perfect battery, i.e., without dissipation, at its stationary state. Hence, in this limit, $\dot{S}_{tot}$ corresponds to the energy dissipation of the circuit only.

In the context of this work, the limit $w_b \to +\infty$ has an analogous meaning: if we consider unidirectional transitions into the system as induced by an external process, characterized by the rates $w_f(t)$ and $w_b$, letting $w_b \to +\infty$ will lead to the correct expression for energy dissipation of the original system{\color{black}, when there are no other sources of dissipation due to external processes}. 

It is important to stress that, {\color{black}within this framework}, the total entropy production is not affected by the limit of an infinite $w_b$, since it depends just on bidirectional transition rates and probabilities (see Eq. \eqref{totalEP}), which are the same in both systems $(A)$ and $(B)$.

We emphasize that the entropy production computed by mapping each unidirectional link into a bidirectional one and letting each fictitious transition rate $w_b\to+\infty$, is equivalent to the direct computation (without mapping) as shown in the Eq. \eqref{parts}, i.e., both of these results are consistent with each other. The aim of introducing such a mapping is to show that the contribution coming only from unidirectional links {\color{black}may have its own physical meaning in some cases. In particular, when there are no external processes dissipating energy, i.e. a perfect battery in the parallel with an electrical circuit, the contribution coming from unidirectional links can be considered as a part of the environment entropy production of the full system. We stress that such a physical identification is crucial to correctly generalize the Schnakenberg's formula.}

%{\color{black}Whether or not the interpretation of $\dot{S}^{(\rm u)}$ as a part of the environment entropy production is always possible (without further corrections) is a question that we leave for future investigations. Indeed, in order to provide a full answer, we need to build a mapping that preserves the correct statistics on each trajectory, and not only on average, as we also point out in the next section.}
{\color{black} The correct interpretation of $\dot{S}^{(\rm u)}$, i.e. whether or not being a part of the environment entropy production (without further corrections), is an interesting question that we leave for future investigations. Indeed, in order to provide a full answer, one needs to build a mapping that preserves the correct statistics on each trajectory (i.e., not only on average level), as we also point out in the next section.}

\subsection{First law with unidirectional transitions}

A thermodynamically consistent choice for the bidirectional transition rates is
\begin{equation}
\frac{w_{ij}}{w_{ji}} = e^{(E_i - E_j)/T},
\label{arrhenius}
\end{equation}
where $k_B  = 1$ for sake of simplicity.

\begin{figure*}[th]\center
    \includegraphics[scale=0.38]{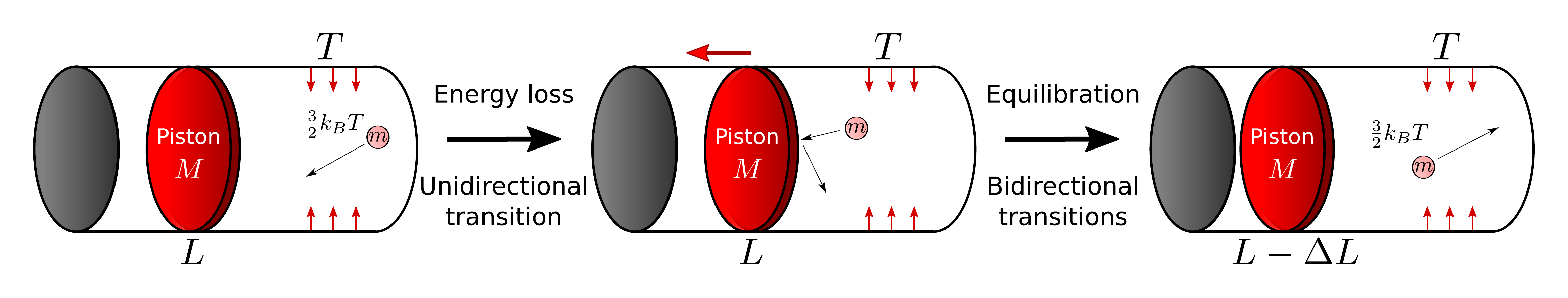}
    \caption{\label{Figpiston} A  physical system with unidirectional and bidirectional transitions. A particle of mass $m$ is contained in a cylinder whose walls are in contact with a heat bath at temperature $T$. A piston of mass $M\gtrsim m$ is placed, and it does not exchange heat with the reservoir. When the particle gets in contact with the wall, it picks a random velocity with energy $E_i$ such that $\langle E_i \rangle= \frac{3}{2} k_B T$ and move ballistically inside the cylinder. Due to its motion, the particle can hit the piston, consequently losing some of its energy, $E_j < E_i$, and moving the piston of a distance $\Delta L$. This event happens with probability $y_{i\to j} dt$, and it can be described as a unidirectional process since the particle cannot receive energy from the piston. After a certain time, the average energy of the particle is restored to $\frac{3}{2} k_B T$, due to collisions with the walls, which are accounted by bidirectional transitions. In this example, the quantity $y_{i\to j} p_i(E_i - E_j)$ is the work done per unit time on the environment.}
\end{figure*}

From Eq.\eqref{dyn-first}, multiplying by $E_i$, summing over $i$, and using Eq.\eqref{arrhenius} for the bidirectional transition rates, it is possible to obtain a formulation of the first law of thermodynamics in presence of unidirectional transitions:
\begin{equation}
\langle \dot{E} \rangle = -T\dot{S}_{env}^{(R)} - \sum_{ij} y_{j \to i} p_j \left( E_j- E_i \right)
\label{firstlaw}
\end{equation}
The first term on the r.h.s. of the previous equation is the heat absorbed by the system from the environment without considering the processes involving unidirectional transitions. The second term, without the minus sign, represents the work extracted from the system, at the expenses of heat absorbed from the thermal bath/environment at temperature $T$, due to events triggered by unidirectional transitions. They both contributes to the total energy change in the system, $\langle \dot{E} \rangle$.
A simple example, which justifies the latter identification is illustrated in Fig.~\ref{Figpiston}. Notice that the term $-\sum_{i,j} y_{j\to i}p_j( \log{p_j}-\log{p_i})$ in Eq.~\eqref{totalEP}, which we identify as a contribution to the environmental entropy production, has a form similar to the second term in Eq.~\eqref{firstlaw} with $E_j-E_i$ substituted by the entropy change $\log{p_j}-\log{p_i}$. This further suggests the correctness of our guess for the environment emtropy production.

\section{Trajectory-based approach}
\label{traj}
In this  section, we will compute the average entropy production using the entropy production along a single stochastic trajectory in a non-equilibrium process. So far we have investigated how the average entropy production can be computed when unidirectional transitions are present in a discrete-state system. The average entropy production can also be explicitly derived using a trajectory-based approach.  This is the standard procedure to derive the well-known fluctuation theorems \cite{seifFT,crooks,jarz2,GCohen}. Here, for completeness, we present in detail this derivation. However, it is important to stress the fact that our framework only preserves average currents, not their fluctuations. As a consequence, the entropy production on a single trajectory may have different statistics between the real and the fictitious system, and only average quantities can be reliably computed through the proposed mapping. We leave for future investigations the problem of defining an entropy production for unidirectional transitions which exhibits the correct statistics on a single trajectory.
 
Consider a system which involves only bidirectional links (see Sec.~\ref{intro}). 
Suppose {\color{black}$\Sigma_{sys}(\tau|\{i(\tau)\})$} be the system entropy along a trajectory {\color{black}$\{i(\tau)\}$} at time $\tau$ (where $0 < \tau < t$), and it is defined as \cite{Andr,seifEP,eft}
\begin{align}
{\color{black}\Sigma_{sys}(\tau|\{i(\tau)\})=-\log p_{i(\tau)}(\tau)}
\label{sys-shanon}
\end{align} 
where {\color{black}$p_{i}(\tau)$ is the solution of the Master equation \eqref{mp} given the initial condition $p_i(0)$}. In the above equation, the trajectory \cite{note} $\{i(\tau)\}=(i_0,i_1,i_2,\dots,i_{M-1},i_M)$, in which $i_k$ is the state of the system from time $\tau_{k}$ to $\tau_{k+1}$, where $\tau_{k}<\tau_{k+1}$, and $\tau_k\in(\tau_0=0,\tau_{M+1}=t)$, and the jumps occur at times $\tau_1,\tau_2,\dots,\tau_M $.

Differentiating the above equation {\color{black}with respect to time}, we find the rate of entropy of the system along a single trajectory 
\begin{widetext}
\begin{align}
\dot\Sigma_{sys}(\tau|\{i(\tau)\}) &= -\dfrac{ \partial_\tau p_{i}(\tau)}{ p_i(\tau)} \bigg|_{i=i(\tau)} -\sum_{k=1}^{M} \delta(\tau-\tau_k) \log \dfrac{ p_{i_k}(\tau_k)}{ p_{i_{k-1}}(\tau_k)}+\delta(\tau_{M+1}-\tau)\log p_{i_M}(\tau_{M+1})-\delta(\tau_{0}-\tau)\log p_{i_0}(\tau_0),
\label{sys-e}
\end{align}
\end{widetext}
where the last two terms are equal to zero as $0=\tau_0<\tau<\tau_{M+1}$.
From the above equation, we can \textcolor{black}{identify} the rate of environment and total entropy as  
\begin{widetext}
\begin{align}
\dot\Sigma_{env}(\tau|\{i(\tau)\}) &=-\sum_{k=1}^{M} \delta(\tau-\tau_k) \log \dfrac{W_{i_{k}\to i_{k-1}}(\tau_k)}{ W_{i_{k-1}\to i_{k}}(\tau_k)},\label{env-e}\\
\dot\Sigma_{tot}(\tau|\{i(\tau)\}) &=-\dfrac{ \partial_\tau p_{i}(\tau)}{ p_i(\tau)} \bigg|_{i=i(\tau)} -\sum_{k=1}^{M} \delta(\tau-\tau_k) \log \dfrac{ p_{i_k}(\tau_k)W_{i_k\to i_{k-1}}(\tau_k)}{ p_{i_{k-1}}(\tau_k)W_{i_{k-1}\to i_{k}}(\tau_k)}\label{tot-e}.
\end{align}
\end{widetext}

In the following, we take the average on \eqref{sys-e}, \eqref{env-e}, and \eqref{tot-e} over trajectories, as defined in \aref{path-prob}. Notice that averaging the first term on the right hand side in \eqref{sys-e}, and \eqref{tot-e}  over $p_i(\tau)$ yields zero due to the normalization $\sum_i p_i(\tau)=1$. Therefore, the rates of average entropy productions are (see \aref{path-prob} for more details)
\begin{align}
\dot S_{sys}&=\sum_{i,j} W_{j\to i}p_j \log\dfrac{p_j}{p_i},\label{eq18}\\
\dot{S}_{env} &= \sum_{i,j} W_{j\to i} p_j\log\dfrac{W_{j\to i}}{W_{i\to j}}\label{eq19},\\
\dot{S}_{tot} &= \sum_{i,j} W_{j\to i} p_j \log\dfrac{W_{j\to i} p_j}{W_{i\to j} p_i}\label{eq20},
\end{align}
where $W$'s are the non-zero transitions.

In the case of environment and total entropy productions, when we consider all the transitions are bidirectional, there is no problem with these quantities. However, as already said before, when some transitions are allowed and the \textcolor{black}{reversed} ones are not, the above $\dot S_{tot}$ and $\dot{S}_{env}$ diverge. Nevertheless, this problem does not arise in the case of system entropy production where one can split $W$ and rewrite the system entropy production as (see Sec. \ref{me-ft}) 
\begin{align}
\dot S_{sys}=\sum_{i,j} w_{j\to i}\ p_j \log\dfrac{p_j}{p_i}+\sum_{i,j} y_{j\to i}\ p_j \log\dfrac{p_j}{p_i},
\end{align}
where, as already said above, the first term involves $w$'s indicating bidirectional transitions whereas the second term involves $y$'s refers to unidirectional transitions. 

In order to deal with this divergence and to give it the right physical interpretation, when some transitions are unidirectional, one can separate those links and replace them by the bidirectional links ($w_f$ and $w_b$ for each unidirectional link) as explained above. Now the system is made by bidirectional links only, then Eqs. \eqref{eq18}, \eqref{eq19} and \eqref{eq20} can be obtained. In the end, after the average over many trajectories, one can take the limit $w_b\to+\infty$ as shown in Sec. \ref{dyna}.

The results obtained following this way, by averaging the trajectory entropy, are consistent with ones derived in the previous sections. In other words, {\color{black}when the limit $w_b \to \infty$ is physically meaningful (in the terms discussed above),} the extra (non-divergent) contribution in $\dot{S}_{sys}$, Eq. \eqref{sys-e}, due to the unidirectional transitions, appears to be equal to (minus) the extra term in the environment entropy production depending just on $y_{i \to j}$. Again, Eq. \eqref{totalEP} holds.

An important observation is that in our framework there is no need of splitting the trajectories into regular and irregular ones \cite{Murashita}. One can first consider the system as completely bidirectional, and then the limit is taken just as a final step. Thus, the average entropy production can also be achieved from each single trajectory and averaging  over them using the procedure explicitly shown in the \aref{path-prob}. 

In the next subsection, we obtain the average entropy production for  an example where some of the transitions are unidirectional.

{\color{black}
\subsection{Partial detailed balance and pseudo-deterministic behaviour}
\label{PDB}
Our framework can be employed to study a particular situation, which we name {\it partial detailed balance}. 

Consider a system in which the detailed balance holds for all the bidirectional links $w_{i \to j}$, i.e., $w_{i \to j} \pi_i = w_{j \to i} \pi_j$, where $\pi_i$ is the probability to be in the state $i$ at stationarity. This condition differs from \textit{global detailed balance}, since unidirectional links do not have to satisfy it. We call it partial detailed balance. 

Let us start with a simple system, as the one depicted in Fig.~\ref{fig-1}. In this case, imposing the detailed balance conditions on all the bidirectional (black) links, the probability of the system to be in the state 3 will evolve according to: 
\begin{equation}
\dot{p}_3(t) = - y_{3 \to 1} p_3(t)
\end{equation}
converges to zero as time progresses. The detailed balance condition then implies that all the probabilities have to be zero at stationarity, which is impossible. Therefore, the detailed balance is never reached on the bidirectional links, since it is inconsistent with the dynamical evolution of the whole system. In other words, the detailed balance is not maintained in such systems in contrast to those systems with only bidirectional links. We emphasize that this argument can be generalized as follows: the partial detailed balance cannot be satisfied for networks where the nodes connected by unidirectional links \textit{do not form cycles}.

\begin{figure*}[th]
\center
\includegraphics[width=1.5 \columnwidth]{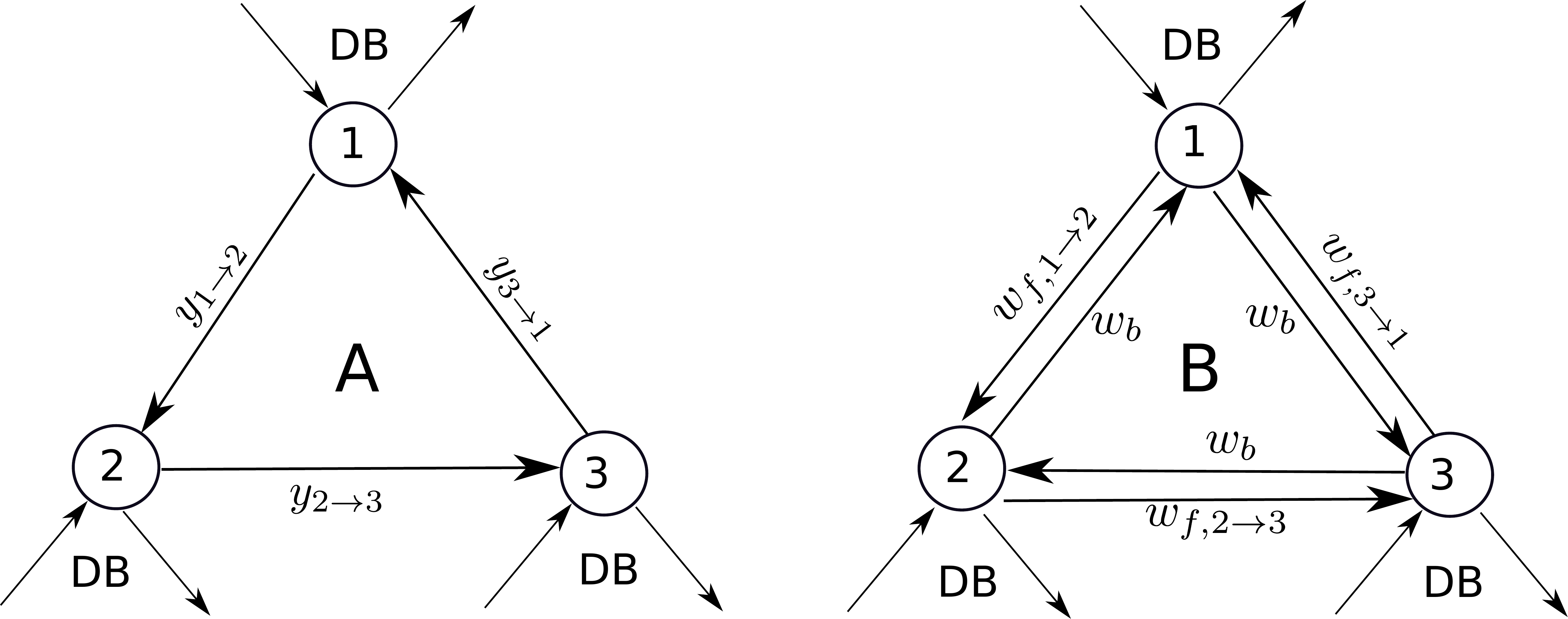}
\caption{\label{three-nodes} A) Three nodes labelled by 1, 2, and 3 are connected by unidirectional links. B) A system equivalent to A is shown where the unidirectional links are replaced by bidirectional links with the transition rates $w_f$'s and $w_b$  as discussed in Eq.~\eqref{fluxes}. Moreover, the nodes 1, 2, and 3 are connected to systems whose transition rates satisfy detailed balance (DB).}
\end{figure*}

We then need to analyze networks in which cycles of unidirectional links do appear. Also for this setting, let us start with the simple example sketched in Fig.~\ref{three-nodes}. Here, when detailed balance holds on bidirectional links, the dynamics still admits a feasible stationary solution for the whole system.

{\color{black}We study this system when the the entropy produced by unidirectional transitions cannot be ascribed to any external process, so that $\dot{S}^{(\rm u)}$ can be identified with an extra contribution to the environment entropy production. As a consequence, the total entropy production is equal to $\dot{S}^{(R)}$, which is the one evaluated taking into account only bidirectional transitions, Eq. \eqref{totalEP}.}

According to this, the total entropy production is equal to zero at stationarity for this particular system, because of the partial detailed balance condition. {\color{black}In other words}, in the presence of cycles of unidirectional links, there can be in principle situations in which the system experiences some transitions without producing entropy, even if global detailed balance does not hold.

In order to unveil the apparent paradox of a null average total entropy production, here we detail how to estimate it starting from trajectory-dependent quantities. Given a trajectory$ \{ i(\tau) \}$, the total entropy change associated to it is 
\begin{equation}
\Delta \Sigma_{tot}(\{ i(\tau) \})=\log\dfrac{\mathcal{P}(\{i(\tau)\})}{\mathcal{P}^\dagger(\{i(\tau)\}^\dagger)} = \int_0^t~d\tau~ \dot{\Sigma}_{tot}(\tau|\{i(\tau)\}) , 
\label{crooks}
\end{equation}
where $\mathcal{P}(\{i(\tau)\})$ is the probability of observing a given trajectory $\{i(\tau)\}$ and $\mathcal{P}^\dagger(\{i(\tau)\}^\dagger)$ is the probability of observing the time reversed trajectory $\{i(\tau)\}^\dagger$. The explicit structure of $\mathcal{P}(\{i(\tau)\})$ is given in  Eq. \ref{px}. A proof of the last equality in Eq. \eqref{crooks} is given in \aref{proof}.

As \textcolor{black}{highlighted} in \cite{rahav}, since we cannot identify a time reversed trajectory for the unidirectional links, we cannot strightforwardly apply Eq.~\eqref{crooks} to identify $\Delta \Sigma_{tot}$. Instead of proposing a different auxiliary dynamics, generating an auxiliary trajectory $\widehat{\{i(\tau)\}}$, which does not correpond to the time-reversed one \cite{rahav}, and would not also lead to the total entropy production on a single trajectory, we start modeling our system with all the transitions as bidirectional links, as explained in Sec.~\ref{dyna}.

The equivalent network is shown in Fig.~\ref{three-nodes}B, where each unidirectional links is replaced by the bidirectional links with the transition rates $w_f$'s and $w_b$ using the procedure presented in Sec. \ref{dyna}. For convenience, we assume $w_b$ to be equal for each reverse transition. Generalization to different $w_b$ is straightforward. In this equivalent network, consider a trajectory $\{ i(\tau) \}$, so that:
\begin{widetext}
\begin{align}
\Delta \Sigma_{tot} (\{ i(\tau) \})=\log \bigg[\dfrac{p_{\text{start}}}{p_{\text{end}}} \bigg(\dfrac{w_{f,1\to2}}{w_{b}}\bigg)^{\mathcal{N}_{1\to 2}-\mathcal{N}_{2\to 1}} \bigg(\dfrac{w_{f,2\to3}}{w_{b}}\bigg)^{\mathcal{N}_{2\to 3}-\mathcal{N}_{3\to 2}} \bigg(\dfrac{w_{f,3\to1}}{w_{b}}\bigg)^{\mathcal{N}_{3\to 1}-\mathcal{N}_{1\to 3}}\bigg],
\label{net-ent}
\end{align}
\end{widetext}
where $\mathcal{N}_{i_k\to i_{k\pm1}}$ is the number of times the transition happened from nodes $i_k$ to $i_{k\pm1}$ in the trajectory $\{i(\tau)\}$. In the above equation, $p_{\text{start}}\  (p_{\text{end}})$ is the probability of the system to be in the initial (final) node from (to) where (which) the trajectory $\{ i(\tau) \}$ starts (ends). In Eq.~\eqref{net-ent}, we have focused only on the cycle formed by unidirectional links. If also the contribution from other nodes connected by bidirectional links, then obeying detailed balance, and belonging to the trajectory $\{ i(\tau) \}$, were considered in Eq. \eqref{net-ent}, it would yield $1$ in the argument of the logarithm by definition. Upon averaging Eq. \eqref{net-ent} over trajectories {\color{black}(which is equivalent to average Eq. \eqref{tot-e} and apply \eqref{crooks})}, then using \eqref{fluxes} and the limit $w_b\to +\infty$, we finally get that the system does not produce entropy on average at stationarity.

An intuitive explanation of this result relies on noting that {\color{black}an intrinsic} unidirectional cycle ({\color{black}not caused by some external process performing work}) exhibits a non-stochastic behaviour. The stochasticity lingers only in the presence of persistence times, which, however, does not contribute to the entropy difference for each trajectory {\color{black}(see Eq. \eqref{net-ent})}, showing what we name a pseudo-deterministic behaviour. Thus, we can conclude that, in the framework {\color{black}here discussed}, unidirectional cycle exhibit null total average entropy production at stationarity. This is a direct consequence of the equivalence derived in Eq. \eqref{totalEP}. In fact, at an average level, bidirectional links are the only ones contributing to the entropy production, even if the $p$'s still depend on the whole network of transitions. 

\section{Application: stochastic resetting}
\label{st-rt}
Stochastic resetting is a mechanism in which the system undergoes a stochastic dynamics in the state space as well as stochastically resets to a prescribed location with a given transition rate (i.e., a unidirectional process) \cite{search1,search2,search3,IFT-arnab,sanjib-relaxation,Arnab-1,pal-time-dep,shamik-apoorva,Uttam,KPZ-power-law,falcao-inter,roldan-path-int,refractory,under-restart,Markov-reset,run-tumble,gupta,gupta2019work,basu2019symmetric}. These resetting transitions involve jumps of the system into given locations and can be called the \textit{controlled transitions} (i.e., these can be tuned from external sources). Notice that corresponding to each controlled transition, there is no reverse transition. Therefore, the whole dynamics has two classes of transition rates: (i) internal or uncontrolled, and (ii) controlled and unidirectional.

\begin{figure*}\center
\includegraphics[width=2 \columnwidth]{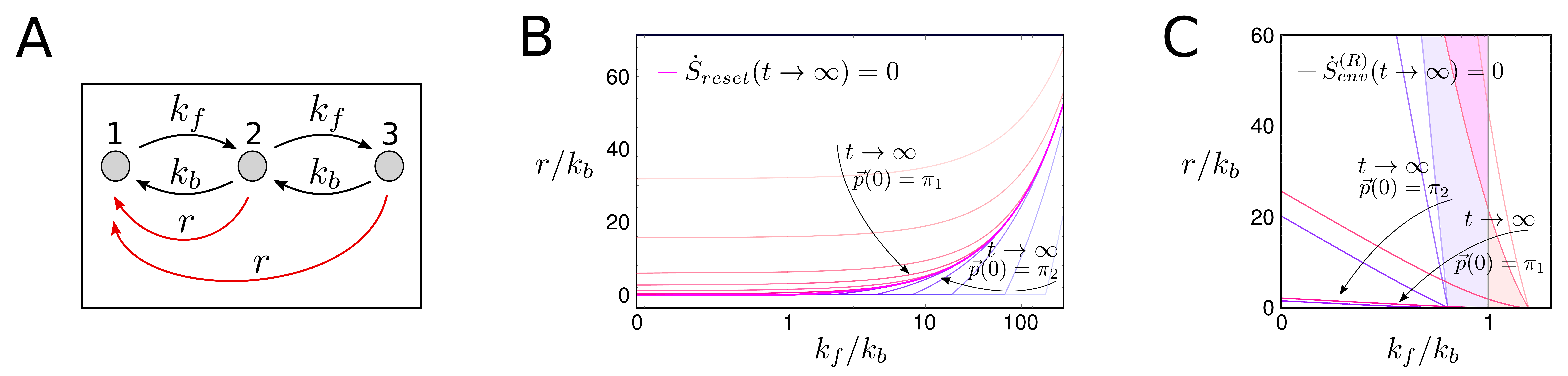}
\caption{\label{fig-4} A) Three-state system with resetting from each state to $1$ at a constant rate $r$. B) The coloured curves $\dot{S}_{reset} \equiv  \dot{S}_{env}^{(R)}-\dot{S}_{env}=0 $, which corresponds to the second term on right hand side in Eq.~\eqref{eqfig}, separate the eraser-like behaviour $\dot S_{reset}<$0 (above the curve) from the writer-like behaviour $\dot S_{reset} >0$ \cite{resetting}. We show how this so-called operative diagram changes as the system approaches stationarity from two different initial conditions: $\pi_1 = (0.15,0.15,0.7)$ and $\pi_2=(0.7,0.15,0.15)$. In general, as a function of time, a resetting system may change its operative behaviour. C) The coloured line, in the same color code of (B), indicates the point for which $\dot{S}_{env}^{(R)}(t)=0$, as the system approaches stationarity from two different initial conditions. $\dot{S}_{env}^{(R)}<0$ from the vertical grey line to the coloured one (coloured region for two times only). The vertical grey line is present for all times and at stationarity divides the region of parameter in which the heat is extracted from the environment ($\dot{S}_{env}^{(R)}<0$ (on the left) from the one in which it is dissipated (on the right)  \cite{resetting}.}
\end{figure*}

Then we can write the entropy production of the system as follows. 
Suppose that a given system resets to a state $i_0$ from a state $j$ with a resetting rate $r_j$. Therefore, using Eq. \eqref{parts}, the system entropy production is
\begin{eqnarray}
\dot S_{sys} &=& \sum_{i,j} w_{j\to i}p_j \log\dfrac{w_{j\to i}p_j}{w_{i\to j}p_i} + \\
&\;& +\overbrace{\sum_{j} r_j p_j \log\dfrac{p_j}{p_{i_0}}}^{\dot S_{reset}}-\sum_{i,j} w_{j\to i}p_j \log\dfrac{w_{j\to i}}{w_{i\to j}}, \nonumber
\label{eqfig}
\end{eqnarray}  
where $w_{i\to j}$ is the transition rate belonging to the uncontrolled process, which we have considered constituted by bidirectional transitions only, and $r_j:=y_{j\to i_0}$ is the resetting rate (where $y_{j\to i_0}y_{i_0\to j}=0$, i.e., controlled process) from the $j$th state. 

In the steady state, $\dot S_{sys}=0$, using Eq. \eqref{totalEP} we are led to:
\begin{equation}
\dot{S}^{(R)}_{tot} =\dot{S}_{tot} \equiv \sum_{i,j} w_{j\to i}p_j \log\dfrac{w_{j\to i}}{w_{i\to j}} - \sum_{j} r_j p_j \log\dfrac{p_j}{p_{i_0}}
\end{equation}
Hence, using our general framework, it is possible to obtain the same result as presented in \cite{resetting}.

{\color{black}In what follows, we identify $\dot{S}^{(\rm u)}$ as $\dot{S}_{reset}$, without further speculating on its physical meaning, since in this case it is univocally determined by the system under investigation.}

\subsection{Constant resetting}
\label{constt-res}
In the following, we consider an example of a network of three nodes labelled by 1, 2, and 3 as shown in Fig.~\ref{fig-4}A. The transition rates from the node 1 to 2 and from 2 to 3 are $k_f$ while the reverse ones are $k_b$. Moreover, we consider that the system stochastically resets from nodes 2 and 3 to node 1 with a rate $r>0$. For this system, we plot the contours for $\dot S_{reset}=0$ (see Fig.~\ref{fig-4}B) and $\dot S^{(R)}_{env}=0$ (see Fig.~\ref{fig-4}C) for two different initial condition $\pi_1 = (0.15,0.15,0.7)$ and $\pi_2=(0.7,0.15,0.15)$ at different times in $(k_f/k_b, r/k_b)$ plane. We observe that as the time increases the system approches the stationary state and the contours $\dot S_{reset}=0$ approaches their stationary value.

Within our framework, we can go beyond the stationary operative diagram, derived in \cite{resetting}. In fact, it is possible to obtain how the region of parameters in which the system exhibits a writer(eraser)-like behaviour changes as time increases. However, this latter is a transient effect, since the system evolves toward a non-equilibrium steady state. It is then worth investigating whether persistent complex phenomena can emerge by introducing a time dependence in the transition rates.

\subsection{Time-periodic resetting}
\label{timeresett}

Time-periodic driving is a non-equilibrium paradigm which can efficiently mimic non-equilibrium features of molecular machines operating at steady state \cite{browne,hern,astumian,raz-subasi,busiello-raz}.  Moreover, several biological systems constantly integrate and produce time-periodic stimuli, to cope with environmental perturbations \cite{bennett, mettetal,tu,timeperiodic}. It is then instructive to study what happens when time-dependent unidirectional transitions come into play. Here, we focus on cases where they implement a time-periodic resetting mechanism.

Consider a system analogous to the one depicted in Fig.~\ref{fig-4}A, where now the resetting rate depends on time through the following rule (as sketched in Fig.~\ref{fig-6}A):
\begin{equation}
r(t) = 2 ~r_0 \cos^2(2 \pi \omega t)
\end{equation}
such that $\langle r(t) \rangle_T = r_0$, where $\langle \cdot \rangle_T$ indicates the temporal average over one period $T = (2\omega)^{-1}$. Both $k_f$ and $k_b$ do not depend on time, so that the periodicity is introduced through the resetting mechanism only. After a transient, the system will relax to a time-periodic state, exhibiting time-periodic environment and resetting entropy production.

For sake of simplicity, we set $k_b = 1$, $k_f = k^+$, and $r$ equal to its average value $r_0$, such that if the system relaxed to a non-equilibrium steady state, the resetting entropy production would be equal to zero, i.e., 
\begin{equation}
\dot{S}_{reset}(t \to \infty) \bigg|_{r = r_0, k_f = k^+, k_b = 1} = 0
\label{Sres0}
\end{equation}
The above equation gives the relation between $r_0$ and $k^+$ corresponding to one point on the magenta curve depicted in Fig.~\ref{fig-4}B. 
%Moreover, we introduce the a reference timescale $\tau$ to span different resetting frequencies in our analysis. 
Since we have different rates acting at the same time on the system, we define a reference timescale $\tau = 1/k_{\rm min}$, where $k_{\min} = \min\{1,k^+,r_0\}$. When the period of the resetting $T=\tau$, the corresponding frequency will be denoted as $\omega_c$, that is, $\omega_c = k_{\rm min}/2$. We then investigate the system for various $\omega$, i.e.,  $\omega<< \omega_c$, $\omega>>\omega_c$ and $\omega = \omega_c$ To this aim, we define $\omega/\omega_c = \alpha $, where $\alpha > 0$ is a dimensionless parameter.

\begin{figure*}\center
\includegraphics[width=2 \columnwidth]{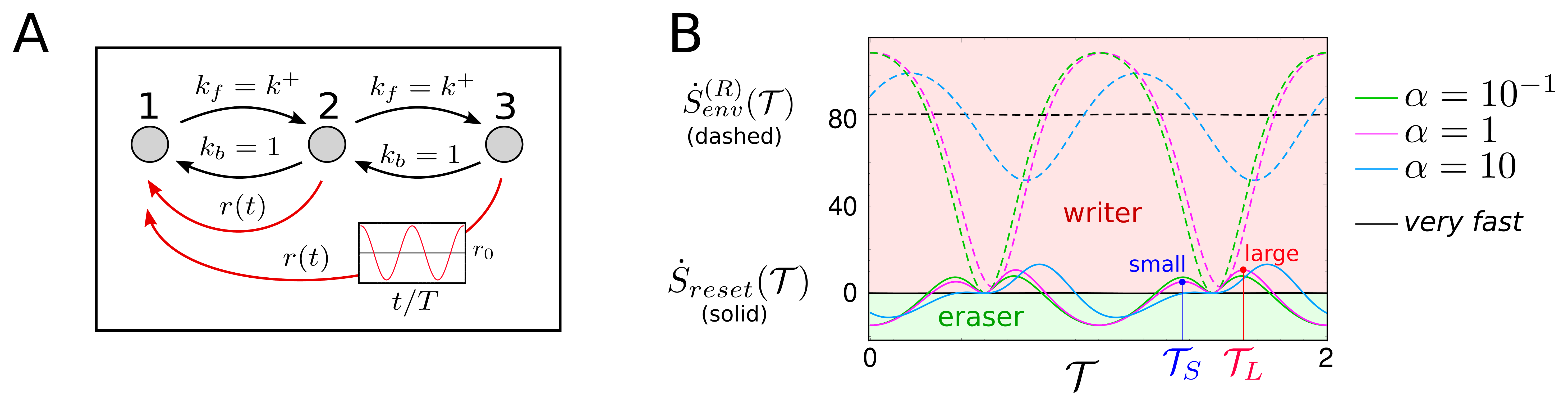}
\caption{\label{fig-6}A) Three-state system with resetting from each state to $1$ at a time dependent rate $r(t) = 2 r_0  \cos^2(2 \pi \omega t)$. We set $k^+ = 40$, so that $r_0=20.579\dots$ from the solution of Eq. \eqref{Sres0}. Here, $\omega_c = k_{\min}/2\equiv 1/2$. B) Solid curves indicate the resetting entropy production as a function of the rescaled time $\mathcal{T}=t/T$, where $t$ is inizialized to $0$ when the system reaches its time-periodic state. Dashed curves represent the environment entropy production. Different colors indicate different values of $\alpha=\omega/\omega_c$, as reported. The system switches between writer-like and eraser-like behaviour (the faster is $\omega$, the faster is the switching). As $\alpha$ increases, the minimum of $\dot{S}^{(R)}_{env}$ increases and the profile of the curve becomes flatter, until being constant for \textit{very fast} driving. On the other hand, in $\dot{S}_{reset}$, an increase in the frequency corresponds to an higher asymmetry between the two positive maxima. The small one is indicated by a blue dot and a blue line, while large maximum by a red dot and a red line, on the curve corresponding to $\alpha = 1$. Note that the small minimum appears before the large in each period, i.e., $\mathcal{T}_S < \mathcal{T}_L$. Further increasing $\alpha$ such an asymmetry starts decreasing, and eventually $\dot{S}_{reset}$ approaches zero at all times for \textit{very fast} drivings, since $r(t)$ can be approximated to its average $r_0$.}
\end{figure*}

\begin{figure*}\center
\includegraphics[width=2 \columnwidth]{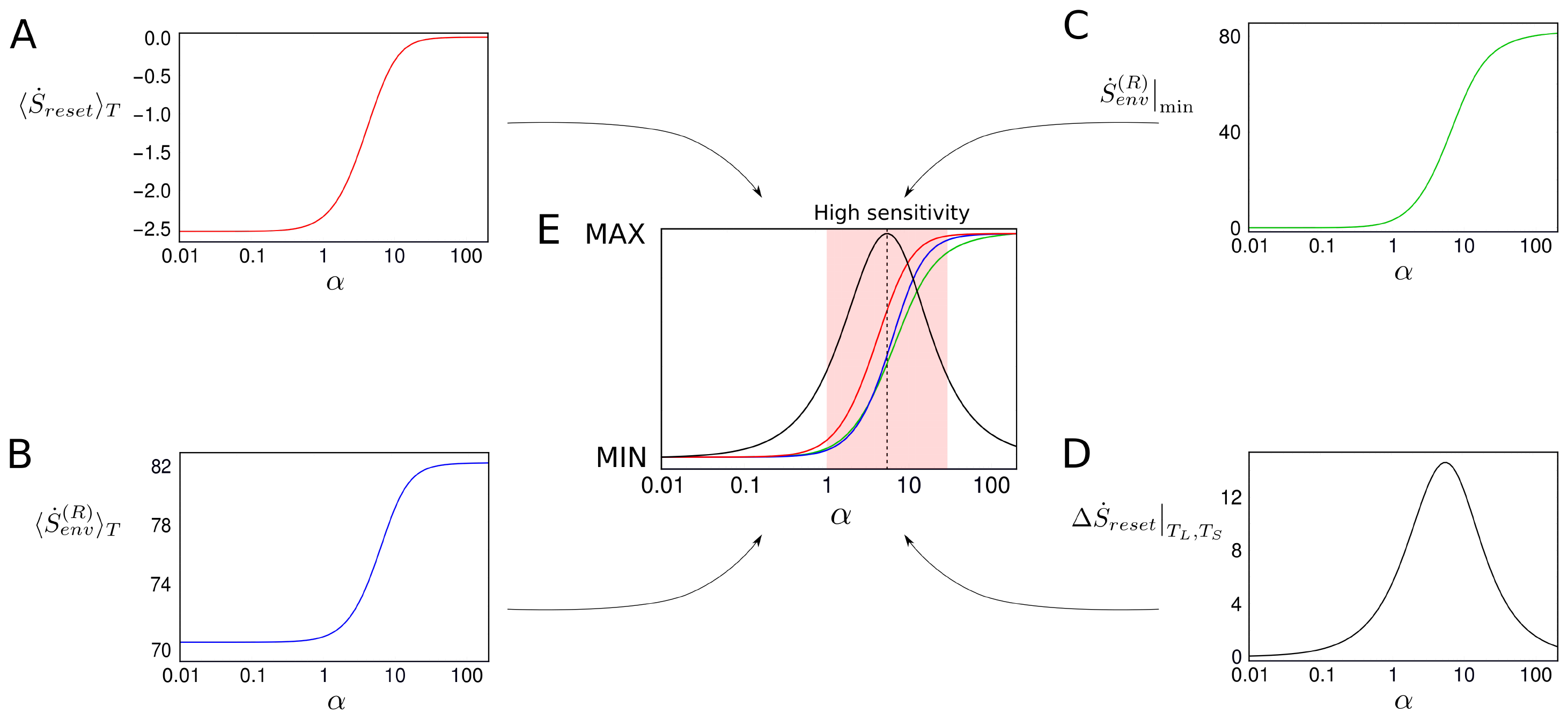}
\caption{\label{fig-7}Several measures of the system in Fig. \ref{fig-6}A are reported. All $x$-axes are in log-scale. A) Temporal average of resetting entropy production over one period. The system, on average, tends to erase information, while $\langle \dot{S}_{reset} \rangle_T \to 0$ for \textit{very fast} drivings, according to Eq. \eqref{Sres0}, since $r(t)$ can be approximated with its average $r_0$. B) Temporal average of environment entropy production. Our framework leads to the thermodynamically consistent result that adiabatic drivings, i.e., $\alpha << 1$, produce less entropy in the surroundings, while the higher is $\alpha$, the higher will be $\langle \dot{S}^{(R)}_{env} \rangle_T$, up to a threshold (constant in time) value. C) Minimum of environment entropy production during one period. Again, in accordance to the law of thermodynamics, the faster is the driving, the larger will be the minimum possible dissipation rate that the system can experience. D) Difference in height of the two positive maxima of $\dot{S}_{reset}$, i.e., $\dot{S}_{reset}(\mathcal{T}_L) - \dot{S}_{reset}(\mathcal{T}_S)$. Both for adiabatic and \textit{very fast} drivings it tends to zero, while presenting a maximum for an intermediate frequency higher than $\omega_c$, the reference frequency of the system. E) Putting together all panels, and normalizing the quantities so to use the same scale on $y$-axis, we sketch, qualitatively, the window of $\alpha$ in which the system is highly sensitive to frequency variations. This region contains mostly frequencies higher than $\omega_c = k_{\min}/2$.}
\end{figure*}

In Fig.~\ref{fig-6} and \ref{fig-7}, we present numerical results. Notice that from now on we study the system for time $t$ such as it has already reached its time-periodic state. Moreover, for sake of simplicity, the time $t$ is considered inizialized to $0$ after the transient dynamics.

First of all, the system exhibits a time-periodic switching between writer-like and eraser-like behavior, even if the resetting entropy produced in the two phases is not the same. Moreover, there are two bumps in the positive region and just one minima in the negative region, in a period $T$. In Fig. \ref{fig-6}B, solid curves represent $\dot{S}_{reset}$ for three different values of $\alpha$ (corresponding to $\omega$ above, below and equal to $\omega_c$). Clearly, as the frequency increases, the speed of the switching increases as well, even if it cannot be seen from the picture, since we are plotting the entropy production against a rescaled time $\mathcal{T} = t/T$. 

Unexpectedly, going toward higher values of $\alpha$, an asymmetry appears in $\dot{S}_{reset}$. In particular, we observe a difference in the heights of the two maxima evidenced by a blue and a red dot at fixed $\alpha$ in the picture. The smaller maximum takes place before the larger one in each period, at (rescaled) time $\mathcal{T}_S<\mathcal{T}_L$, where $\mathcal{T}_L$ is the (rescaled) time at which the larger maximum appears. As $\alpha$ is further increased, such an asymmetry disappers again, and eventually the resetting entropy production vanishes at all times. This behavior is a consequence of the fact that, for \textit{very fast} drivings, $r(t)$ can be approximated with its average $r_0$, for which $\dot{S}_{reset}$ is zero by construction, Eq. \eqref{Sres0}.

Remarkably, this observation can be rephrased in a more fashonable way: time-periodic resetting introduces a signature of the arrow of time. It is known that entropy production is a time-antisymmetric quantity whose average over many dissipative trajectories can help in discriminating if a process is proceeding forward or backward, i.e., to identify the arrow of time \cite{parrondo}. Here we discuss how time-dependent unidirectional links (performing a resetting to a given state) can lead, in principle, to an error-free discrimination between forward and backward processes. In fact, just by looking at the order in which the maxima appear in $\dot{S}_{reset}$, during a period, we can conclude: if the smaller maximum comes first, then time is proceeding, viceversa, time is regressing. However, the bottleneck is represented by the long-standing problem about how to measure entropy production \cite{horo}. Moreover, we are reporting results from a simple example rather than a general theory, and we leave for future works a more in-depth analysis of this preliminary - and peculiar - observation.

Dashed lines in Fig. \ref{fig-6}B represent the environment entropy production due to the internal transitions. It can be seen that both its average and its minimum value increase when $\alpha$ increases at fixed $\mathcal{T}$. Eventually it approaches the stationary value at all times for \textit{very fast} drivings. A thermodynamic argument can justify this result, since adiabatic (\textit{slow}) protocols, in general, can attain lower level of dissipation in the surroundings. However, such a consistency between thermodynamics and our results serves as a further endorsement of the framework here introduced.

We introduce four measures to quantify this phenomenology, and to study the system as a function of frequency of $r(t)$:
\begin{gather}
\label{resint}
\langle \dot{S}_{reset}(t) \rangle_T = \frac{1}{T} \int_T \dot{S}_{reset}(t)~ dt \\
\label{envint}
\langle \dot{S}^{(R)}_{env}(t) \rangle_T = \frac{1}{T} \int_T \dot{S}^{(R)}_{env}(t)~ dt \\
\label{envmin}
\dot{S}^{(R)}_{env} \big|_{\min} \equiv \min_{t \in \{T\}} \dot{S}^{(R)}_{env}(t) \\
\label{diffmax}
\Delta \dot{S}_{reset} \big|_{\mathcal{T}_L,\mathcal{T}_S} \equiv \dot{S}_{reset}(\mathcal{T}_L) - \dot{S}_{reset}(\mathcal{T}_S)
\end{gather}

Eq. \eqref{resint} quantifies the temporal average of the entropy production due to the resetting mechanism over one period $T$. In Fig.~\ref{fig-7}A we show that it is always negative, and approaches zero for large $\alpha$. This means that on average the system erases information.

Eq. \eqref{envint} is the temporal average of the entropy production in the environment over one period, and, as reported in Fig.~\ref{fig-7}B, it increases with $\alpha$, consistently with thermodynamics and our previous qualitative observation. An analogous behaviour is displayed by the minimum amount of dissipation in the surroundings, defined in Eq. \eqref{envmin}, and shown in Fig.~\ref{fig-7}C.

As noticed above, the asymmetry between the positive maxima of $\dot{S}_{reset}$ increases with $\alpha$ up to a maximum value, and then goes back to $0$ for \textit{very fast} drivings (see Fig. \ref{fig-7}D).

Putting together all these informations, we produce a qualitative graph in Fig. \ref{fig-7}E. Here all the quantities have been normalized so to be seen on the same scale. It appears natural to identify a finite window of $\alpha$ in which the system exhibits high sensitivity to the variation of the resetting frequency. In fact, the slope of $\langle \dot{S}_{reset}(t) \rangle_T, \langle \dot{S}^{(R)}_{env}(t) \rangle_T$ and $\dot{S}^{(R)}_{env} \big|_{\min}$ is maximized in the this region, and it is also (roughly) centered around the maximum of $\Delta \dot{S}_{reset} \big|_{\mathcal{T}_L,\mathcal{T}_S}$. It is worth noting that this high sensitivity window contains mostly frequencies higher than the reference time-scale of the system, associated to the minimum average rate $k_{\min}$. However, we stress the fact that this study provides just a qualitative analysis of an example, and it does not aim at being exhaustive about the effect of time-periodic unidirectional links. It is true that a rich and complex phenomenology arises even for the simple system analyzed, and we believe that this topic deserves much attention and leaves several questions for future studies.

\section{Conclusions}
\label{conc}
In this work we have proposed a theoretical (and physically meaningful) argument to generalize the Schnakenberg's entropy production to cases in which unidirectional transitions are allowed. This kind of situation emerges when several different processes are in play to drive the system out of equilibrium, and it is also relevant for various physical and biological applications.

The solution we have adopted has the advantage of preserving the dynamics of the system at all times, and it can also be derived from a trajectory-based approach. Moreover, it provides a direct procedure to deal with the presence of unidirectional transitions.

We have studied the peculiar case in which unidirectional cycles are present, deriving that, from a thermodynamic perspective, they behave like pseudo-deterministic systems.

Within our framework, we have studied the resetting problem, which is one of the most prominent applications in which unidirectional transitions are involved. We have shown that previous results in this context can thus be recovered in a broader perspective. Moreover, we have generalized the analysis to time-periodic resetting, finding that the system can switch between an eraser-like and a writer-like behaviour. Moreover, our framework is consistent with the fact that the faster is the driving, the higher is the entropy production in the environment. On top of this, fast drivings lead to a signature of the arrow of time in the resetting entropy production, an observation which can be a starting point for future investigations.

The presented framework can be nevertheless applied to several different situations, from population dynamics to biological modelizations, leaving a wide panorama of possibilities for further applications.

\section*{Acknowledgements}
D. Gupta and A. Maritan acknowledge the support from University of Padova  through ``Excellence Project 2018'' of the Cariparo foundation.

%%%%%%%%%%%%%%%%%%%%%%%%%%%%%%%%%%%%%%%%%%%%%%%%%%%%%%%%%%%%%%%%%%%%%%%%%%%%
%%%%%%%%%%%%%%%%%%%%%%%%%%%%%%%%%%%%%%%%%%%%%%%%%%%%%%%%%%%%%%%%%%%%%%%
%% This part is added (a modification in the iopart.cls) to modify the
%% tableofcontents entry for the appendices
%%%%%%%%%%%%%%%%%%%%%%%%%%%%%%%%%%%%%%%%%%%%%%%%%%%%%%%%%%%%%%%%%%%%%%%
%\makeatletter%
%\def\@sect#1#2#3#4#5#6[#7]#8{\ifnum #2>\c@secnumdepth
 % \let\@svsec\@empty\else
 % \refstepcounter{#1}\edef\@svsec{\csname the#1\endcsname. }\fi
  %\@tempskipa #5\relax
  %\ifdim \@tempskipa>\z@
  %\begingroup #6\relax
  %\noindent{\hskip #3\relax\@svsec}{\interlinepenalty \@M #8\par}%
  %\endgroup
  %\csname #1mark\endcsname{#7}\addcontentsline
   %       {toc}{#1}{\ifnum #2>\c@secnumdepth \else
    %        \protect\numberline{}{\csname the#1\endcsname}\fi
     %       . #7}\else
      %    \def\@svsechd{#6\hskip #3\relax  %% \relax added 2 May 90
       %     \@svsec #8\csname #1mark\endcsname
        %            {#7}\addcontentsline
         %           {toc}{#1}{\ifnum #2>\c@secnumdepth \else
          %            \protect\numberline{}{\csname the#1\endcsname}\fi
           %           . #7}}\fi
  %        \@xsect{#5}}
%\makeatother%
%%%%%%%%%%%%%%%%%%%%%%%%%%%%%%%%%%%%%%%%%%%%%%%%%%%%%%%%%%%%%%%%%

%\newpage
\appendix
%\begin{widetext}
\section{Average entropy production}
\label{path-prob}
In this section, we compute the average entropy production starting from the entropy production given for a single stochastic trajectory. 

Consider a system which evolves according to the Master Equation in Eq. \eqref{mp}. Given a function of the trajectory, $F(\{i(\tau)\})$ (e.g. $\int_0^t g(i(\tau)) d\tau$ with $g$ a generic function defined on the ensemble of states), we define the average over trajectories as 
\begin{equation}
\langle F \rangle_{\rm traj} = \sum_{M=0}^{+\infty} \sum^{(M)}_{i(\tau)} {\color{black}\mathcal{P}(\{i(\tau)\})} F(\{i(\tau)\}),
\end{equation}
where we introduced the following summation:
\begin{equation}
\sum^{(M)}_{i(\tau)} \equiv \sum_{\substack{i_{M},i_{M-1},\\i_{M-2}, \dots,i_1,i_0}} \prod_{l=1}^{M} \bigg(\int_0^t\ d\tau_l\bigg) \prod_{j=1}^{M+1}\Theta(\tau_j-\tau_{j-1}),
\label{sum}
\end{equation}
where $\Theta(\cdot)$ is the Heaviside-theta function.  This summation runs over all possible trajectories involving $M$ jumps and lasting a time $t$. Then, ${\color{black}\mathcal{P}(\{i(\tau)\})}$ is the probability of observing such a trajectory:
\begin{align}
{\color{black}\mathcal{P}(\{i(\tau)\})} =&~ Q_i(t,\tau_M)W_{i_{M-1}\to {\color{black}i_M}}(\tau_M)\times\nonumber\\ &~Q_{i_{M-1}}(\tau_{M},\tau_{M-1}) \dots  W_{i_{0}\to i_1}(\tau_1)\times\nonumber\\& ~Q_{i_0}(\tau_1,0)p_{i_0}(0).
\label{px}
\end{align}
{\color{black}Notice that $p_j(\tau)$, the solution of the master equation can be obtained as $p_j(\tau)=\langle \delta_{j,i(\tau)}\rangle_{\mathrm{traj}}$.

In the above equation, $p_{i_0}(0)$ is the initial distribution, and 
\begin{align}
Q_{i_X}(t,t')=\exp\bigg[-\int_{t^\prime}^t dt_1\ \mathcal{W}_{i_X}(t_1) \bigg],
\end{align}
where $\mathcal{W}_{i_X}(t)=\sum_{j_X} W_{i_X\to j_X}$ is the transition rate of the exit from state $i_X$.}

We are interested in the case $F = \dot{\Sigma}_{sys}$, defined in Eq. \eqref{sys-e}. Thus, the average entropy production of the system over all the trajectories is given as 
{\color{black}\begin{align}
\dot S_{sys}(\tau)=\langle \dot{\Sigma}_{sys} \rangle=&\sum_{M=0}^{+\infty} \sum^{(M)}_{i(\tau)} \mathcal{P}(\{i(\tau)\}) \sum_{k=1}^{M} \delta(\tau-\tau_k)\times \nonumber\\
 &f(i_k,i_{k-1},\tau_k),
\end{align}}
where $f(i_k,i_{k-1},\tau_k)=\log \dfrac{ p_{i_{k-1}}(\tau_k)}{ p_{i_{k}}(\tau_k)}$. We stress that averaging the first term on the right hand side of Eq. \eqref{sys-e} we get:
{\color{black}\begin{align}
\left\langle \frac{\partial_\tau p_i(\tau)}{p_i(\tau)}\bigg|_{i=i(\tau)} \right\rangle_{\rm traj} =\left\langle \sum_i \frac{\partial_\tau p_i(\tau)}{p_i(\tau)} \delta_{i(\tau),i} \right\rangle_{\rm traj}  \nonumber\\= \sum_i \frac{\partial_\tau p_i(\tau)}{p_i(\tau)} \langle \delta_{i(\tau),i} \rangle_{\rm traj}  \nonumber\\= \sum_i \partial_\tau p_i(\tau) = 0.
\end{align}}
In the last but one step we have used the fact that {\color{black}$\partial_\tau p_i(\tau)/p_i(\tau)$} does not depend anymore on the trajectory.

%The probability distribution function of the system \eqref{mp} to be at $x$ at time $t$ starting from $x_0$ at time $t_0=0$  is given as
%\begin{align}
%P(x,t)=\sum_{M=0}^\infty \sum_{\substack{x_{M-1},x_{M-2},\\x_{M-3}, \dots,x_1,x_0}} \prod_{i=1}^{M} \bigg(\int_0^t\ d\tau_i\bigg) \prod_{j=1}^{M+1}\Theta(\tau_j-\tau_{j-1})\mathcal{P}_M(\{x(\tau)\}),
%\label{px}
%\end{align}
%where {\color{black}$\Theta(x)$} is the Heaviside-theta function, and $\mathcal{P}_{M}(\{x(\tau)\})$ is the probability of a trajectory over $M$ states in a time $t$:
%\begin{align}
%\mathcal{P}_{M}(\{x(\tau)\})=Q_x(t,\tau_M)W_{x_{M-1}\to x}(\tau_M)\ Q_{x_{M-1}}(\tau_{M},\tau_{M-1}) \dots  W_{x_{0}\to x_1}(\tau_1)\ Q_{x_0}(\tau_1,0)P(x_0,0).
%\label{ptraj}
%\end{align}
%Note that $x$ represents a discrete state.

%In the above equation, $P(x_0,0)$ is the initial distribution, and 
%\begin{align}
%Q_x(t,t')=\exp\bigg[-\int_{t^\prime}^t dt_1\ \mathcal{W}_x(t_1) \bigg],
%\end{align}
%where $\mathcal{W}_x(t)=\sum_{y} W_{x\to y}$ is the transition rate of the exit from state $x$. 

Using Eqs. \eqref{sum} and \eqref{px}, we can explicitly compute this average:
\begin{align}
{\color{black}\dot S_{sys}(\tau)}=&\sum_{M=0}^\infty \sum_{\substack{i_M,i_{M-1},..\\ \dots,i_1,i_0}} \sum_{k=1}^{M}\prod_{l=1}^{M} \bigg(\int_0^t\ d\tau_l\bigg)\times\nonumber\\
  & \prod_{j=1}^{M+1}\Theta(\tau_j-\tau_{j-1}) Q_{i_M}(t,\tau_M)W_{i_{M-1}\to i_M}(\tau_M)\times\nonumber\\
&  Q_{i_{M-1}}(\tau_{M},\tau_{M-1}) W_{i_{M-2}\to i_{M-1}}(\tau_{M-1})\dots \times \nonumber\\ & Q_{i_{k}}(\tau_{k+1},\tau_{k}) W_{i_{k-1}\to i_{k}}(\tau_k)  \delta(\tau-\tau_k)\times  \nonumber\\ 
& f(i_k,i_{k-1},\tau_k) Q_{i_{k-1}}(\tau_{k},\tau_{k-1})\times \nonumber\\   &W_{i_{k-2}\to i_{k-1}}(\tau_{k-1})\dots  W_{i_{0}\to i_1}(\tau_1)\times\nonumber\\
& Q_{i_0}(\tau_1,0)p_{i_0}(0),
\end{align}
where $i=i_M$ and $t=\tau_{M+1}$.

We first note that
\begin{equation}
\sum_{M=0}^{+\infty} \sum_{k=1}^M \Rightarrow \sum_{k=1}^{+\infty} \sum_{M=k}^{+\infty}.
\end{equation}
Since $k$ has to be smaller or equal than $M$, $M$ can go up to $\infty$, and the term $k = 0$ does not contribute to the average entropy production.

We can rewrite the expression above as follows:
\begin{widetext}
\begin{align}
{\color{black}\dot S_{sys}(\tau)}&=\sum_{k=1}^\infty \sum_{\substack{i_{k},\dots,i_0}} \left\{ \sum_{i_M} \left[\sum_{M=k}^{+\infty} \sum_{i_{M-1},\dots,i_{k+1}} \prod_{l=k+1}^{M} \bigg(\int_0^t\ d\tau_l\bigg) \prod_{j=k+1}^{M+1}\Theta(\tau_j-\tau_{j-1}) Q_{i_M}(t,\tau_M)  \right. \right. \times  \nonumber\\
&W_{i_{M-1}\to i_M}(\tau_M)\ Q_{i_{M-1}}(\tau_{M},\tau_{M-1}) W_{i_{M-2}\to i_{M-1}}(\tau_{M-1})\dots Q_{i_{k}}(\tau_{k+1},\tau_{k}) \bigg] \bigg\}\times \nonumber\\
&  \prod_{l=1}^{k-1} \bigg(\int_0^t\ d\tau_l\bigg) \prod_{j=1}^{k}\Theta(\tau_j-\tau_{j-1}) W_{i_{k-1}\to i_k}(\tau)\ Q_{i_{k-1}}(t,\tau_{k-1})\times \nonumber\\
&  W_{i_{k-2}\to i_{k-1}}(\tau_{k-1})\dots f(i_k,i_{k-1},\tau) \dots W_{i_{0}\to i_1}(\tau_1)\ Q_{i_0}(\tau_1,0)p_{i_0}(0).
\end{align}
\end{widetext}
In the above equation, we have performed the integral over $\tau_k$ while using $\delta(\tau-\tau_k)$. {\color{black}In the product of $\Theta(\tau_j-\tau_{j-1})$ it is implicitly assumed that $\tau_k=\tau$. } Now, the term in the square bracket can be identified as $p(i_M,t|i_k,\tau)$, so that:
\begin{widetext}
\begin{align}
{\color{black}\dot S_{sys}(\tau)}&=\sum_{k=1}^\infty \sum_{\substack{i_{k},\dots,i_0}} \prod_{l=1}^{k-1} \bigg(\int_0^t\ d\tau_l\bigg) \prod_{j=1}^{k}\Theta(\tau_j-\tau_{j-1}) \left\{ \sum_{i_M} p(i_M,t|i_k,\tau) \right\} W_{i_{k-1}\to i_k}(\tau)\times\nonumber\\
&  Q_{i_{k-1}}(\tau,\tau_{k-1}) W_{i_{k-2}\to i_{k-1}}(\tau_{k-1})\dots f(i_k,i_{k-1},\tau) \dots W_{i_{0}\to i_1}(\tau_1)\ Q_{i_0}(\tau_1,0)p_{i_0}(0),
\end{align}
\end{widetext}
and, the term in the curly brackets is equal to $1$. Thus, renaming $\kappa = k-1$, we obtain:
\begin{widetext}
\begin{align}
\dot S_{sys}(\tau) &= \sum_{\kappa=0}^\infty \sum_{i_{\kappa + 1},i_{\kappa}} \bigg[ \sum_{\substack{i_{\kappa-1},\dots,i_0}} \prod_{l=1}^{\kappa} \bigg(\int_0^t\ d\tau_l\bigg) \prod_{j=1}^{\kappa + 1}\Theta(\tau_j-\tau_{j-1}) Q_{i_{\kappa}}(\tau,\tau_{\kappa}) W_{i_{\kappa-1}\to i_{\kappa}}(\tau_{\kappa})\times \nonumber\\
& W_{i_{0}\to i_1}(\tau_1)\ Q_{i_0}(\tau_1,0)p_{i_0}(0) \bigg] W_{i_{\kappa}\to i_{\kappa + 1}}(\tau)\ f(i_{\kappa+1},i_{\kappa},\tau).
\end{align}
\end{widetext}
The term in the square brackets is nothing but $p_{i_\kappa}(\tau)$, then, naming $i = i_{\kappa + 1}$ and $j = i_{\kappa}$, which are any two connected states, and substituting the expression of {\color{black}$f(i_{\kappa+1},i_{\kappa},\tau)$}, we finally get:
\begin{align}
{\color{black}\dot S_{sys}(\tau)}&=\sum_{i,j} W_{j\to i}(\tau)\ p_j(\tau) \log\dfrac{p_j(\tau)}{p_i(\tau)},
\end{align}
{\color{black} which is nothing else that  Eq. \eqref{eq18}.}

Similarly, one can do the averaging over environment and total entropy given in \eqref{env-e} and \eqref{tot-e}, respectively, to obtain respective the entropy productions {\color{black} given in Eqs. \eqref{eq19} and \eqref{eq20}}.

\section{Relation between \eqref{tot-e} and \eqref{crooks} }
\label{proof}
In this section, we aim to find the relation between \eqref{tot-e} and \eqref{crooks}. {\color{black}Therefore, we consider \eqref{tot-e} 
\begin{align}
&\dot{\Sigma}_{tot}(\tau|\{i(\tau)\})=-\dfrac{ \partial_\tau p_{i}(\tau)}{ p_i(\tau)} \bigg|_{i=i(\tau)}+\nonumber\\
 &~~~~~~~~-\sum_{l=1}^{M} \delta(\tau-\tau_l) \log \dfrac{ p_{i_l}(\tau_l)W_{i_l\to i_{l-1}}(\tau_l)}{ p_{i_{l-1}}(\tau_l)W_{i_{l-1}\to i_{l}}(\tau_l)}.
\end{align}

Now we rewrite the above equation as
\begin{align}
&\dot{\Sigma}_{tot}(\tau|\{i(\tau)\})=\dot{\sigma}_1+\dot{\sigma}_2,
\end{align}
where
\begin{align}
\dot{\sigma}_1&=-\dfrac{\partial_\tau p_{i}(\tau)}{p_i(\tau)} \bigg|_{i=i(\tau)}-\sum_{l=1}^{M} \delta(\tau-\tau_l) \log \dfrac{p_{i_l}(\tau_l)}{p_{i_{l-1}}(\tau_l)},\nonumber\\
\dot{\sigma}_2&=\sum_{l=1}^{M} \delta(\tau-\tau_l) \log \dfrac{ p_{i_l}(\tau_l)}{ p_{i_{l-1}}(\tau_l)}+\nonumber\\&~~~~~ -\sum_{l=1}^{M} \delta(\tau-\tau_l) \log \dfrac{ p_{i_l}(\tau_l)W_{i_l\to i_{l-1}}(\tau_l)}{ p_{i_{l-1}}(\tau_l)W_{i_{l-1}\to i_{l}}(\tau_l)}.
\end{align}
Now $\dot{\sigma}_1$ is the total derivative of $\Sigma_{sys}(\tau|\{i(\tau)\}) = -\log{p_{i(\tau)}}(\tau)$, as shown in Eqs. \eqref{sys-shanon} and \eqref{sys-e}. We integrate the above equation over time $\tau$ from 0 to {\color{black}$t$ with $i(0)=i_0$ and $i(t)=i$} and obtain
\begin{align}
\sigma_1&=-\log \dfrac{p_i(t)}{p_{i_0}(0)},\\
\sigma_2&=\log\prod_{l=1}^M\dfrac{p_{i_l}(\tau_l)}{p_{i_{l-1}}(\tau_l)}+\nonumber\\
&-\log\prod_{l=1}^M\dfrac{p_{i_l}(\tau_l)W_{i_l\to i_{l-1}}(\tau_l)}{p_{i_{l-1}}(\tau_l)W_{i_{l-1}\to i_{l}}(\tau_l)}\nonumber\\
& \equiv\log\prod_{l=1}^M\dfrac{W_{i_{l-1}\to i_{l}}(\tau_l)}{W_{i_{l}\to i_{l-1}}(\tau_l)}.
\end{align}

Combining $\sigma_1$ and $\sigma_2$ yields the total entropy production \eqref{crooks} along a stochastic trajectory.
}
%\end{widetext}

\bigskip

%\bibliographystyle{acm}
%\bibliography{draft.bib}

\begin{thebibliography}{}


\bibitem{schn}
{\sc Schnakenberg, J.}
\newblock Network theory of microscopic and macroscopic behavior of master
  equation systems.
\newblock {\em Reviews of Modern physics 48}, 4 (1976), 571.







\bibitem{barato}
{\sc Barato, A.~C., and Seifert, U.}
\newblock Cost and precision of brownian clocks.
\newblock {\em Physical Review X 6}, 4 (2016), 041053.



\bibitem{dechant}
{\sc Dechant, A., and Sasa, S.-i.}
\newblock Entropic bounds on currents in langevin systems.
\newblock {\em Physical Review E 97}, 6 (2018), 062101.



\bibitem{Esposito-CG}
{\sc Esposito, M.}
\newblock Stochastic thermodynamics under coarse graining.
\newblock {\em Phys. Rev. E 85\/} (Apr 2012), 041125.

\bibitem{jarz}
{\sc Floyd, C., Jarzynski, C., and Papoian, G.~A.}
\newblock Quantifying dissipation in actomyosin networks.
\newblock {\em Biophysical Journal 116}, 3 (2019), 254a.



\bibitem{horo}
{\sc Li, J., Horowitz, J.~M., Gingrich, T.~R., and Fakhri, N.}
\newblock Quantifying dissipation using fluctuating currents.
\newblock {\em Nature communications 10}, 1 (2019), 1666.


\bibitem{baiesi}
{\sc Baiesi, M., Maes, C., and Wynants, B.}
\newblock Fluctuations and response of nonequilibrium states.
\newblock {\em Physical review letters 103}, 1 (2009), 010602.




\bibitem{prigogine}
{\sc Prigogine, I.}
\newblock {\'E}tude thermodynamique des ph{\'e}nom{\`e}nes irr{\'e}versibles.



\bibitem{vanden}
{\sc Jiu-Li, L., Van Den~Broeck, C., and Nicolis, G.}
\newblock Stability criteria and fluctuations around nonequilibrium states.
\newblock {\em Zeitschrift f{\"u}r Physik B Condensed Matter 56}, 2 (1984),
  165--170.



\bibitem{Ronald}
{\sc Eduardo~de Oliveira~Rodrigues, J., and Dickman, R.}
\newblock Asymmetric exclusion process in a system of interacting brownian
  particles.
\newblock {\em Phys. Rev. E 81\/} (Jun 2010), 061108.


\bibitem{saha}
{\sc Saha, B., and Mukherji, S.}
\newblock Entropy production and large deviation function for systems with
  microscopically irreversible transitions.
\newblock {\em Journal of Statistical Mechanics: Theory and Experiment 2016}, 1
  (2016), 013202.





\bibitem{Ting}
{\sc Ting, C.~L., Makarov, D.~E., and Wang, Z.-G.}
\newblock A kinetic model for the enzymatic action of cellulase.
\newblock {\em The Journal of Physical Chemistry B 113}, 14 (2009), 4970--4977.
\newblock PMID: 19292431.




\bibitem{Chong}
{\sc Chong, S.-H., Otsuki, M., and Hayakawa, H.}
\newblock Generalized {G}reen-{K}ubo relation and integral fluctuation theorem
  for driven dissipative systems without microscopic time reversibility.
\newblock {\em Phys. Rev. E 81\/} (Apr 2010), 041130.



\bibitem{Masaki}
{\sc Takeuchi, K.~A., Kuroda, M., Chat\'e, H., and Sano, M.}
\newblock Directed percolation criticality in turbulent liquid crystals.
\newblock {\em Phys. Rev. Lett. 99\/} (Dec 2007), 234503.



\bibitem{rahav}
{\sc Rahav, S., and Harbola, U.}
\newblock An integral fluctuation theorem for systems with unidirectional
  transitions.
\newblock {\em Journal of Statistical Mechanics: Theory and Experiment 2014},
  10 (2014), P10044.



\bibitem{search1}
{\sc Evans, M.~R., and Majumdar, S.~N.}
\newblock Diffusion with stochastic resetting.
\newblock {\em Physical review letters 106}, 16 (2011), 160601.

{\color{black}
\bibitem{reviewresetting}
{\sc Evans, M.~R., Majumdar, S.~N., and Schehr, G.}
\newblock Stochastic resetting with applications.
\newblock {\em arXiv preprint arXiv:1910.07993\/} (2019)}


\bibitem{Paolo}
{\sc Assenza, S., Sassi, A.~S., Kellner, R., Schuler, B., Rios, P. D.~L., and
  Barducci, A.}
\newblock Efficient conversion of chemical energy into mechanical work by hsp70
  chaperones.
\newblock {\em arXiv preprint arXiv:1902.01612\/} (2019).



\bibitem{hinric}
{\sc Zeraati, S., Jafarpour, F.~H., and Hinrichsen, H.}
\newblock Entropy production of nonequilibrium steady states with irreversible
  transitions.
\newblock {\em Journal of Statistical Mechanics: Theory and Experiment 2012},
  12 (2012), L12001.




\bibitem{ben2011entropy}
{\sc fBen Avraham, D., Dorosz, S., and Pleimling, M.}
\newblock Entropy production in nonequilibrium steady states: A different
  approach and an exactly solvable canonical model.
\newblock {\em Physical Review E 84}, 1 (2011), 011115.

\bibitem{Murashita}
{\sc Murashita, Y., Funo, K., and Ueda, M.}
\newblock Nonequilibrium equalities in absolutely irreversible processes.
\newblock {\em Phys. Rev. E 90\/} (Oct 2014), 042110.


\bibitem{ohkubo}
{\sc Ohkubo, J.}
\newblock Posterior probability and fluctuation theorem in stochastic
  processes.
\newblock {\em Journal of the Physical Society of Japan 78}, 12 (2009),
  123001--123001.

\bibitem{lacasta}
{\sc Lacasta, A., Sancho, J., Romero, A., and Lindenberg, K.}
\newblock Sorting on periodic surfaces.
\newblock {\em Physical review letters 94}, 16 (2005), 160601.


\bibitem{Andr}
{\sc Andrieux, D., and Gaspard, P.}
\newblock Fluctuation theorem for currents and schnakenberg network theory.
\newblock {\em Journal of statistical physics 127}, 1 (2007), 107--131.



\bibitem{search2}
{\sc Meylahn, J.~M., Sabhapandit, S., and Touchette, H.}
\newblock Large deviations for markov processes with resetting.
\newblock {\em Physical Review E 92}, 6 (2015), 062148.

\bibitem{search3}
{\sc Kusmierz, L., Majumdar, S.~N., Sabhapandit, S., and Schehr, G.}
\newblock First order transition for the optimal search time of l{\'e}vy
  flights with resetting.
\newblock {\em Physical review letters 113}, 22 (2014), 220602.


\bibitem{proof1}
{\sc Murugan, A., Huse, D.~A., and Leibler, S.}
\newblock Discriminatory proofreading regimes in nonequilibrium systems.
\newblock {\em Physical Review X 4}, 2 (2014), 021016.

\bibitem{proof2}
{\sc Hartich, D., Barato, A.~C., and Seifert, U.}
\newblock Nonequilibrium sensing and its analogy to kinetic proofreading.
\newblock {\em New Journal of Physics 17}, 5 (2015), 055026.

\bibitem{popdyn}
{\sc Dharmaraja, S., Di~Crescenzo, A., Giorno, V., and Nobile, A.~G.}
\newblock A continuous-time {E}hrenfest model with catastrophes and its
  jump-diffusion approximation.
\newblock {\em Journal of Statistical Physics 161}, 2 (2015), 326--345.

\bibitem{info}
{\sc Parrondo, J.~M., Horowitz, J.~M., and Sagawa, T.}
\newblock Thermodynamics of information.
\newblock {\em Nature physics 11}, 2 (2015), 131.


\bibitem{resetting}
{\sc Fuchs, J., Goldt, S., and Seifert, U.}
\newblock Stochastic thermodynamics of resetting.
\newblock {\em EPL (Europhysics Letters) 113}, 6 (2016), 60009.


\bibitem{browne}
{\sc Browne, W.~R., and Feringa, B.~L.}
\newblock Making molecular machines work.
\newblock In {\em Nanoscience and Technology: A Collection of Reviews from
  Nature Journals}. World Scientific, 2010, pp.~79--89.


\bibitem{hern}
{\sc Hern{\'a}ndez, J.~V., Kay, E.~R., and Leigh, D.~A.}
\newblock A reversible synthetic rotary molecular motor.
\newblock {\em Science 306}, 5701 (2004), 1532--1537.


\bibitem{busiello-raz}
{\sc Busiello, D.~M., Jarzynski, C., and Raz, O.}
\newblock Similarities and differences between non-equilibrium steady states
  and time-periodic driving in diffusive systems.
\newblock {\em New Journal of Physics 20}, 9 (2018), 093015.

\bibitem{astumian}
{\sc Astumian, R.~D.}
\newblock Adiabatic operation of a molecular machine.
\newblock {\em Proceedings of the National Academy of Sciences 104}, 50 (2007),
  19715--19718.


\bibitem{raz-subasi}
{\sc Raz, O., Suba\ifmmode \mbox{\c{s}}\else \c{s}\fi{}\ifmmode \imath \else~\i
  \fi{}, Y., and Jarzynski, C.}
\newblock Mimicking nonequilibrium steady states with time-periodic driving.
\newblock {\em Phys. Rev. X 6\/} (May 2016), 021022.


\bibitem{bennett}
{\sc Bennett, M.~R., Pang, W.~L., Ostroff, N.~A., Baumgartner, B.~L., Nayak,
  S., Tsimring, L.~S., and Hasty, J.}
\newblock Metabolic gene regulation in a dynamically changing environment.
\newblock {\em Nature 454}, 7208 (2008), 1119.


\bibitem{timeperiodic}
{\sc de~Ronde, W.~H., Tostevin, F., and Ten~Wolde, P.~R.}
\newblock Effect of feedback on the fidelity of information transmission of
  time-varying signals.
\newblock {\em Physical Review E 82}, 3 (2010), 031914.


\bibitem{mettetal}
{\sc Mettetal, J.~T., Muzzey, D., G{\'o}mez-Uribe, C., and van Oudenaarden, A.}
\newblock The frequency dependence of osmo-adaptation in saccharomyces
  cerevisiae.
\newblock {\em Science 319}, 5862 (2008), 482--484.


\bibitem{tu}
{\sc Tu, Y., Shimizu, T.~S., and Berg, H.~C.}
\newblock Modeling the chemotactic response of escherichia coli to time-varying
  stimuli.
\newblock {\em Proceedings of the National Academy of Sciences 105}, 39 (2008),
  14855--14860.



\bibitem{busielloPRE}
{\sc Busiello, D.~M., Hidalgo, J., and Maritan, A.}
\newblock Entropy production in systems with random transition rates close to
  equilibrium.
\newblock {\em Physical Review E 96}, 6 (2017), 062110.






\bibitem{seifFT}
{\sc Seifert, U.}
\newblock Stochastic thermodynamics, fluctuation theorems and molecular
  machines.
\newblock {\em Reports on progress in physics 75}, 12 (2012), 126001.



\bibitem{jarz2}
{\sc Jarzynski, C.}
\newblock Nonequilibrium equality for free energy differences.
\newblock {\em Physical Review Letters 78}, 14 (1997), 2690.


\bibitem{GCohen}
{\sc Gallavotti, G., and Cohen, E.~G.}
\newblock Dynamical ensembles in nonequilibrium statistical mechanics.
\newblock {\em Physical review letters 74}, 14 (1995), 2694.



\bibitem{crooks}
{\sc Crooks, G.~E.}
\newblock Entropy production fluctuation theorem and the nonequilibrium work
  relation for free energy differences.
\newblock {\em Physical Review E 60}, 3 (1999), 2721.











\bibitem{eft}
{\sc Esposito, M., Harbola, U., and Mukamel, S.}
\newblock Entropy fluctuation theorems in driven open systems: Application to
  electron counting statistics.
\newblock {\em Phys. Rev. E 76\/} (Sep 2007), 031132.


\bibitem{seifEP}
{\sc Seifert, U.}
\newblock Entropy production along a stochastic trajectory and an integral
  fluctuation theorem.
\newblock {\em Physical review letters 95}, 4 (2005), 040602.

\bibitem{note}
\newblock $i(\tau) = \sum_{k+0}^M i_k \left[ \Theta\left(\tau_{k+1} - \tau \right) - \Theta\left(\tau_k - \tau \right) \right]$, and $\log p_{i(\tau)}(\tau) = \sum_{k+0}^M \log p_{i_k}(\tau) \left[ \Theta\left(\tau_{k+1} - \tau \right) - \Theta\left(\tau_k - \tau \right) \right]$, where $\Theta(\cdot)$ is the Heaviside theta function.



\bibitem{basu2019symmetric}
{\sc Basu, U., Kundu, A., and Pal, A.}
\newblock Symmetric exclusion process under stochastic resetting.
\newblock {\em Physical Review E 100}, 3 (2019), 032136.


\bibitem{Uttam}
{\sc Bhat, U., Bacco, C.~D., and Redner, S.}
\newblock Stochastic search with poisson and deterministic resetting.
\newblock {\em Journal of Statistical Mechanics: Theory and Experiment 2016}, 8
  (2016), 083401.









\bibitem{run-tumble}
{\sc Evans, M.~R., and Majumdar, S.~N.}
\newblock Run and tumble particle under resetting: a renewal approach.
\newblock {\em Journal of Physics A: Mathematical and Theoretical 51}, 47
  (2018), 475003.

\bibitem{refractory}
{\sc Evans, M.~R., and Majumdar, S.~N.}
\newblock Effects of refractory period on stochastic resetting.
\newblock {\em Journal of Physics A: Mathematical and Theoretical 52}, 1
  (2019), 01LT01.

\bibitem{falcao-inter}
{\sc Falcao, R., and Evans, M.~R.}
\newblock Interacting brownian motion with resetting.
\newblock {\em Journal of Statistical Mechanics: Theory and Experiment 2017}, 2
  (2017), 023204.



\bibitem{gupta}
{\sc Gupta, D.}
\newblock Stochastic resetting in underdamped brownian motion.
\newblock {\em Journal of Statistical Mechanics: Theory and Experiment 2019}, 3
  (mar 2019), 033212.

\bibitem{gupta2019work}
{\sc Gupta, D., Plata, C.~A., and Pal, A.}
\newblock Work fluctuations and jarzynski equality in stochastic resetting.
\newblock {\em arXiv preprint arXiv:1909.08512\/} (2019).

\bibitem{KPZ-power-law}
{\sc Gupta, S., and Nagar, A.}
\newblock Resetting of fluctuating interfaces at power-law times.
\newblock {\em Journal of Physics A: Mathematical and Theoretical 49}, 44
  (2016), 445001.





\bibitem{sanjib-relaxation}
{\sc Majumdar, S.~N., Sabhapandit, S., and Schehr, G.}
\newblock Dynamical transition in the temporal relaxation of stochastic
  processes under resetting.
\newblock {\em Phys. Rev. E 91\/} (May 2015), 052131.


\bibitem{Markov-reset}
{\sc Meylahn, J.~M., Sabhapandit, S., and Touchette, H.}
\newblock Large deviations for markov processes with resetting.
\newblock {\em Phys. Rev. E 92\/} (Dec 2015), 062148.


\bibitem{shamik-apoorva}
{\sc Nagar, A., and Gupta, S.}
\newblock Diffusion with stochastic resetting at power-law times.
\newblock {\em Phys. Rev. E 93\/} (Jun 2016), 060102.


\bibitem{Arnab-1}
{\sc Pal, A.}
\newblock Diffusion in a potential landscape with stochastic resetting.
\newblock {\em Phys. Rev. E 91\/} (Jan 2015), 012113.

\bibitem{pal-time-dep}
{\sc Pal, A., Kundu, A., and Evans, M.~R.}
\newblock Diffusion under time-dependent resetting.
\newblock {\em Journal of Physics A: Mathematical and Theoretical 49}, 22
  (2016), 225001.

\bibitem{IFT-arnab}
{\sc Pal, A., and Rahav, S.}
\newblock Integral fluctuation theorems for stochastic resetting systems.
\newblock {\em Phys. Rev. E 96\/} (Dec 2017), 062135.

\bibitem{under-restart}
{\sc Pal, A., and Reuveni, S.}
\newblock First passage under restart.
\newblock {\em Phys. Rev. Lett. 118\/} (Jan 2017), 030603.




\bibitem{roldan-path-int}
{\sc Rold\'an, E., and Gupta, S.}
\newblock Path-integral formalism for stochastic resetting: Exactly solved
  examples and shortcuts to confinement.
\newblock {\em Phys. Rev. E 96\/} (Aug 2017), 022130.



\bibitem{parrondo}
{\sc Parrondo, J.~M., Van~den Broeck, C., and Kawai, R.}
\newblock Entropy production and the arrow of time.
\newblock {\em New Journal of Physics 11}, 7 (2009), 073008.



\end{thebibliography}
{}

\end{document}